%% file: main.tex
\documentclass[
  journal=pasa,
  manuscript=research-paper, %% or "review"
  year=2022,
  volume=37,
]{cup-journal}

\usepackage{microtype,siunitx,booktabs}
\usepackage{graphicx}
\usepackage{amsmath,amssymb}
\usepackage{pdflscape}

\usepackage{hyperref} 
\hypersetup{colorlinks,citecolor=blue,linkcolor=blue,urlcolor=blue}

\title{WALLABY Pilot Survey: The diversity of HI structural parameters in nearby galaxies}

\author{T.N.~Reynolds}
\affiliation{International Centre for Radio Astronomy Research (ICRAR), The University of Western Australia, 35 Stirling Hwy, Crawley, WA, 6009, Australia}
\alsoaffiliation{ARC Centre of Excellence for All Sky Astrophysics in 3 Dimensions (ASTRO 3D), Australia}
\email[T. N. Reynolds]{tristan.reynolds@uwa.edu.au}

\author{B.~Catinella}
\affiliation{International Centre for Radio Astronomy Research (ICRAR), The University of Western Australia, 35 Stirling Hwy, Crawley, WA, 6009, Australia}
\alsoaffiliation{ARC Centre of Excellence for All Sky Astrophysics in 3 Dimensions (ASTRO 3D), Australia}

\author{L.~Cortese}
\affiliation{International Centre for Radio Astronomy Research (ICRAR), The University of Western Australia, 35 Stirling Hwy, Crawley, WA, 6009, Australia}
\alsoaffiliation{ARC Centre of Excellence for All Sky Astrophysics in 3 Dimensions (ASTRO 3D), Australia}

\author{N.~Deg}
\affiliation{Department of Physics, Engineering Physics, and Astronomy, Queen's University, Kingston ON K7L 3N6, Canada}

\author{H.~D\'{e}nes}
\affiliation{ASTRON - The Netherlands Institute for Radio Astronomy, 7991 PD Dwingeloo, The Netherlands}

\author{A.~Elagali}
\affiliation{School of Biological Sciences, The University of Western Australia, Perth, WA, Australia}
\alsoaffiliation{Minderoo Foundation, Perth, WA, Australia}

\author{B.-Q.~For}
\affiliation{International Centre for Radio Astronomy Research (ICRAR), The University of Western Australia, 35 Stirling Hwy, Crawley, WA, 6009, Australia}
\alsoaffiliation{ARC Centre of Excellence for All Sky Astrophysics in 3 Dimensions (ASTRO 3D), Australia}

\author{P.~Kamphuis}
\affiliation{Ruhr University Bochum, Faculty of Physics and Astronomy, Astronomical Institute (AIRUB), D-44780 Bochum, Germany}

\author{D.~Kleiner}
\affiliation{INAF – Osservatorio Astronomico di Cagliari, Via della Scienza 5, 09047 Selargius, CA, Italy}

\author{B.S.~Koribalski}
\affiliation{CSIRO Space and Astronomy, PO Box 76, Epping, NSW 1710 Australia}
\alsoaffiliation{School of Science, Western Sydney University, Locked Bag 1797, Penrith, NSW 2751, Australia}

\author{K.~Lee-Waddell}
\affiliation{International Centre for Radio Astronomy Research (ICRAR), The University of Western Australia, 35 Stirling Hwy, Crawley, WA, 6009, Australia}
\alsoaffiliation{CSIRO Space and Astronomy, PO Box 1130, Bentley, WA 6102, Australia}

\author{C.~Murugeshan}
\affiliation{CSIRO Space and Astronomy, PO Box 1130, Bentley, WA 6102, Australia}
\alsoaffiliation{ARC Centre of Excellence for All Sky Astrophysics in 3 Dimensions (ASTRO 3D), Australia}

\author{W.~Raja}
\affiliation{CSIRO Space and Astronomy, PO Box 76, Epping, NSW 1710 Australia}

\author{J.~Rhee}
\affiliation{International Centre for Radio Astronomy Research (ICRAR), The University of Western Australia, 35 Stirling Hwy, Crawley, WA, 6009, Australia}
\alsoaffiliation{ARC Centre of Excellence for All Sky Astrophysics in 3 Dimensions (ASTRO 3D), Australia}

\author{K.~Spekkens}
\affiliation{Department of Physics and Space Science, Royal Military College of Canada, PO Box 17000, Station Forces, Kingston, Ontario, Canada, K7L 2E1}

\author{L.~Staveley-Smith}
\affiliation{International Centre for Radio Astronomy Research (ICRAR), The University of Western Australia, 35 Stirling Hwy, Crawley, WA, 6009, Australia}
\alsoaffiliation{ARC Centre of Excellence for All Sky Astrophysics in 3 Dimensions (ASTRO 3D), Australia}

\author{J.M.~van~der~Hulst}
\affiliation{Kapteyn Astronomical Institute, University of Groningen, Landleven 12, 9747AD Groningen, The Netherlands}

\author{J.~Wang}
\affiliation{Kavli Institute for Astronomy and Astrophysics, Peking University, Beijing 100871, China}

\author{T.~Westmeier}
\affiliation{International Centre for Radio Astronomy Research (ICRAR), The University of Western Australia, 35 Stirling Hwy, Crawley, WA, 6009, Australia}
\alsoaffiliation{ARC Centre of Excellence for All Sky Astrophysics in 3 Dimensions (ASTRO 3D), Australia}

\author{O.I.~Wong}
\affiliation{CSIRO Space and Astronomy, PO Box 1130, Bentley, WA 6102, Australia}
\alsoaffiliation{International Centre for Radio Astronomy Research (ICRAR), The University of Western Australia, 35 Stirling Hwy, Crawley, WA, 6009, Australia}
\alsoaffiliation{ARC Centre of Excellence for All Sky Astrophysics in 3 Dimensions (ASTRO 3D), Australia}

\author{F.~Bigiel}
\affiliation{Argelander-Institut f\"ur Astronomie, Universit\"at Bonn, Auf dem H\"ugel 71, 53121 Bonn, Germany}

\author{A.~Bosma}
\affiliation{Aix Marseille Universit\'e, CNRS, CNES, LAM, Marseille, France}

\author{B.W.~Holwerda}
\affiliation{Department of physics and astronomy, University of Louisville, Natural Science Building 102, 40292 KY, Louisville, USA}

\author{D.A.~Leahy}
\affiliation{Department of Physics and Astronomy, University of Calgary, Calgary, AB, Canada, T2N 1N4}

\author{M.J.~Meyer}
\affiliation{International Centre for Radio Astronomy Research (ICRAR), The University of Western Australia, 35 Stirling Hwy, Crawley, WA, 6009, Australia}
\alsoaffiliation{ARC Centre of Excellence for All Sky Astrophysics in 3 Dimensions (ASTRO 3D), Australia}

\doi{10.1017/pasa.2022.xxx}

\received {dd Mmm YYYY}
\revised  {dd Mmm YYYY}
\accepted {dd Mmm YYYY}
\published{dd Mmm YYYY}

 %% First letter not capped
\begin{document}

\input{content}

% PASA uses footnotes, not endnotes. \endnote in this template will behave like \footnote; and \printendnotes will not output anything.
% \printendnotes

\bibliography{master}

\appendix

\input{appendices}

\end{document}

%% file: content.tex
\begin{abstract}
We investigate the diversity in the sizes and average surface densities of the neutral atomic hydrogen (H\,\textsc{i}) gas discs in $\sim280$ nearby galaxies detected by the Widefield ASKAP L-band Legacy All-sky Blind Survey (WALLABY). We combine the uniformly observed, interferometric H\,\textsc{i} data from pilot observations of the Hydra cluster and NGC\,4636 group fields with photometry measured from ultraviolet, optical and near-infrared imaging surveys to investigate the interplay between stellar structure, star formation and H\,\textsc{i} structural parameters. We quantify the H\,\textsc{i} structure by the size of the H\,\textsc{i} relative to the optical disc and the average H\,\textsc{i} surface density measured using effective and isodensity radii. For galaxies resolved by $>1.3$ beams, we find that galaxies with higher stellar masses and stellar surface densities tend to have less extended H\,\textsc{i} discs and lower H\,\textsc{i} surface densities: the isodensity H\,\textsc{i} structural parameters show a weak negative dependence on stellar mass and stellar mass surface density. These trends strengthen when we limit our sample to galaxies resolved by $>2$ beams. We find that galaxies with higher H\,\textsc{i} surface densities and more extended H\,\textsc{i} discs tend to be more star forming: the isodensity H\,\textsc{i} structural parameters have stronger correlations with star formation. Normalising the H\,\textsc{i} disc size by the optical effective radius (instead of the isophotal radius) produces positive correlations with stellar masses and stellar surface densities and removes the correlations with star formation. This is due to the effective and isodensity H\,\textsc{i} radii increasing with mass at similar rates while, in the optical, the effective radius increases slower than the isophotal radius. Our results are in qualitative agreement with previous studies and demonstrate that with WALLABY we can begin to bridge the gap between small galaxy samples with high spatial resolution H\,\textsc{i} data and large, statistical studies using spatially unresolved, single-dish data.
\end{abstract}

\section{INTRODUCTION}
\label{sec:intro}

Neutral atomic hydrogen (H\,\textsc{i}) plays an integral role in galaxy evolution as it provides the raw material for future star formation via conversion to molecular hydrogen (H$_2$). The conversion from atomic to molecular gas occurs once the atomic phase reaches sufficiently high densities ($\sim10\,\mathrm{M}_{\odot}\,\mathrm{pc}^{-2}$, e.g.\ \citeauthor{Martin2001} \citeyear{Martin2001}; \citeauthor{Bigiel2008} \citeyear{Bigiel2008}; \citeauthor{Leroy2008} \citeyear{Leroy2008}; \citeauthor{Bigiel2010} \citeyear{Bigiel2010}), hence how the H\,\textsc{i} is distributed throughout a galaxy will affect its ability to form H$_2$ and ultimately stars (e.g.\ there is little star formation observed beyond the optical disc where H\,\textsc{i} tends to be at low densities, $\lesssim2\,\mathrm{M}_{\odot}\,\mathrm{pc}^{-2}$, e.g.\ \citeauthor{Bigiel2010} \citeyear{Bigiel2010}; \citeauthor{Wang2014} \citeyear{Wang2014}). This highlights the importance of understanding the interplay between the distribution of H\,\textsc{i} in galaxies and other galaxy quantities (e.g.\ stellar structure and star formation) and in particular the H\,\textsc{i} that is co-located with regions of star formation (i.e.\ within the optical disc). 

Measuring the size of a galaxy's stellar, star forming or H\,\textsc{i} disc is not a straightforward endeavour as there is no well defined edge to a galaxy. Deeper and more sensitive observations continue to find emission from various galaxy components at increasing radii \citep[e.g.][]{Meurer2018}. As a result, astronomers have used a variety of different definitions when measuring the size of galaxies \citep[for discussion and comparison of different size definitions see e.g.][]{Cortese2012, MunozMateos2015, Sanchez-Almeida2020, Trujillo2020}. Two common definitions of size used for optical images are the radius encompassing a fraction of the total light (e.g.\ 50\% or 90\%, $R_{50}$ and $R_{90}$, respectively, e.g.\ \citeauthor{devaucouleurs1948} \citeyear{devaucouleurs1948}) and the radius at which the surface brightness reaches a set isophotal limit (e.g.\ 25\,mag\,arcsec$^{-2}$, $R_{\rm{iso}}$, e.g.\ \citeauthor{Redman1936} \citeyear{Redman1936}; \citeauthor{Liller1960} \citeyear{Liller1960}). Each size definition has its advantages and disadvantages. The effective (50\% total light) radius is easy to measure and is less dependent on the depth of the observations \citep[e.g.][]{Trujillo2001}, however it is very sensitive to the brightness distribution \citep[e.g.\ the presence of a bright bulge,][]{Trujillo2020}. The isophotal radius is insensitive to the presence of a bulge and provides a better measure of the total size of a disc galaxy, but has the drawback of being limited by the observation depth (e.g.\ 23.5--26.5\,mag\,arcsec$^{-2}$, \citeauthor{Holmberg1958} \citeyear{Holmberg1958}; \citeauthor{Lauberts1989} \citeyear{Lauberts1989}; \citeauthor{Hall2012} \citeyear{Hall2012}). Isophotal radii also produce a tighter stellar size-mass relation than effective radii and better trace the total stellar mass \citep[e.g.][]{Saintonge2011, Cortese2012, MunozMateos2015, Sanchez-Almeida2020, Trujillo2020}.

Unlike optical disc radii, which are frequently defined using both isophotal and fraction of total light enclosed, H\,\textsc{i} discs are normally defined using an isodensity radius measured at $1\,\mathrm{M}_{\odot}\,\mathrm{pc}^{-2}$ (i.e.\ the typical H\,\textsc{i} column density sensitivity reached in observations, e.g.\ \citeauthor{Wang2016} \citeyear{Wang2016}). Studies of spatially resolved H\,\textsc{i} discs find a very tight relation between a galaxy's H\,\textsc{i} mass, $M_{\rm{HI}}$, and isodensity diameter, $D_{\rm{iso,HI}}$ \citep[e.g.][]{Broeils1997, Swaters2002, Wang2016, Rajohnson2022}. Using $\sim550$ galaxies from a compilation of interferometric H\,\textsc{i} surveys, \cite{Wang2016} derive a best fit H\,\textsc{i} size-mass relation of
\begin{equation}
    \log(D_{\mathrm{iso,HI}}/\mathrm{kpc}) = 0.506 \log(M_{\mathrm{HI}}/\mathrm{M}_{\odot}) - 3.293 
\end{equation}
with a scatter of $\sigma=0.06$\,dex. This relation holds over $\sim5$\,dex in H\,\textsc{i} mass and suggests that all galaxies (i.e.\ both low and high mass) have almost identical H\,\textsc{i} radial surface density profiles \citep[e.g.][]{Wang2016} and that the different galaxy types have similar average H\,\textsc{i} surface densities \citep{Broeils1997}. \cite{Stevens2019b} used simulated, mock H\,\textsc{i} galaxies to show that this relation is a product of the shape of H\,\textsc{i} radial surface density profiles, which tend to follow an approximately exponential shape. \cite{Stevens2019b} also measured the H\,\textsc{i} size-mass relation using effective radii for their mock galaxies and find that the scatter of the relation increases relative to their isodensity size-mass relation.

However, the assumed universality of H\,\textsc{i} surface density profile shapes does not hold towards the centres of the radial H\,\textsc{i} distribution. In The H\,\textsc{i} Nearby Galaxy Survey \citep[THINGS,][]{Walter2008}, the radial H\,\textsc{i} surface densities within the central $\sim6$--12\,kpc ($\sim0.5R_{\rm{iso}}$) vary by $\sim1$--$10\,\mathrm{M}_{\odot}\,\mathrm{pc}^{-2}$ in different galaxies \citep{Leroy2008, Bigiel2010}. Similar variations in the central H\,\textsc{i} radial surface density profiles are also found in the Bluedisk survey \citep{Wang2014} and the Westerbork H\,\textsc{i} Survey of Spiral and Irregular Galaxies \citep[WHISP,][]{Swaters2002}. Variations in the radial H\,\textsc{i} surface density profiles of galaxies are found to correlate with various galaxy quantities including galaxy morphological type \citep[e.g.][]{Cayatte1994, Broeils1997}, H$_2$ and star formation rate surface densities \citep[e.g.][]{Leroy2008, Bigiel2010}, stellar mass and $\mathrm{NUV}-r$ colour \citep[e.g.][]{Wang2014}. However, the interferometric surveys used in these studies contain only modest sample sizes (e.g.\ THINGS: 23, Bluedisk: 50, WHISP: 73).

Investigating how galaxy properties regulate the H\,\textsc{i} gas content using large galaxy samples has been limited to single-dish surveys (e.g.\ the Arecibo Legacy Fast ALFA survey -- ALFALFA, \citeauthor{Giovanelli2005} \citeyear{Giovanelli2005}, and the extended \textit{GALEX} Arecibo SDSS Survey -- xGASS, \citeauthor{Catinella2018} \citeyear{Catinella2018}; see \citeauthor{Saintonge2022} \citeyear{Saintonge2022} for a review). Using xGASS, \cite{Catinella2018} find that the total H\,\textsc{i} gas fraction is more strongly correlated with star formation (e.g.\ specific star formation rate and $\mathrm{NUV}-r$ colour) than with stellar mass and stellar mass surface density. However, studying the total H\,\textsc{i} gas fraction does not provide information on where the H\,\textsc{i} gas is located (e.g.\ within or beyond the optical disc).

Taking advantage of the tight relation between H\,\textsc{i} size and mass, \cite{Wang2020} developed a method of estimating the H\,\textsc{i} gas mass within the optical disc from spatially unresolved, single-dish observations. \cite{Wang2020} find that the scatter in the H\,\textsc{i} gas fraction scaling relations derived for H\,\textsc{i} detected disc galaxies in xGASS decreases if the total H\,\textsc{i} mass is replaced with the estimated H\,\textsc{i} mass within the optical disc. This result was also found by \cite{Naluminsa2021} using a sample of 228 WHISP galaxies. Using the \cite{Wang2020} method, \cite{Chen2022} find a stronger correlation between the gas-phase metallicity and the H\,\textsc{i} mass within the optical disc than with the total H\,\textsc{i} mass. \cite{Pan2021} also used the tightness of the H\,\textsc{i} size-mass relation to estimate H\,\textsc{i} sizes from integrated H\,\textsc{i} masses and find a weak trend of less extended H\,\textsc{i} discs relative to their optical discs in more massive galaxies.

The previous works discussed here have taken one of two avenues to investigate the spatial distribution of H\,\textsc{i} in galaxies. Many studies have looked at the spatial H\,\textsc{i} distribution in detail for small samples of galaxies with spatially resolved H\,\textsc{i} data \citep[$\gtrsim10$ beams along the galaxy's major axis, e.g.][]{Cayatte1994, Broeils1997, Swaters2002, Noordermeer2005, Leroy2008, Wang2013, Wang2014, Wang2016, Wang2017}. A smaller number have taken a statistical approach and estimated simple quantities parameterising the H\,\textsc{i} distribution from large single-dish surveys \citep[e.g.][]{Wang2020, Pan2021, Chen2022}. Recent and upcoming surveys on new radio interferometers such as the Australian Square Kilometre Array Pathfinder \citep[ASKAP,][]{Johnston2008, Hotan2021}, the Karoo Array Telescope \citep[MeerKAT,][]{Jonas2016} and the APERture Tile in Focus upgrade on the Westerbork Synthesis Telescope \citep[APERTIF,][]{Verheijen2008, Adams2022} will bridge this gap and enable direct measurements quantifying the H\,\textsc{i} spatial distribution (i.e.\ the radial H\,\textsc{i} surface density distribution) for statistically significant samples of galaxies.

The Widefield ASKAP L-band Legacy All-sky Blind Survey \citep[WALLABY,][]{Koribalski2020} is one such survey providing this capability. WALLABY is currently underway on ASKAP and will detect H\,\textsc{i} emission in $\sim210\,000$ galaxies and spatially resolve several thousand galaxies over $\sim1.4\pi$\,sr of the southern sky out to $z\sim0.1$ \citep{Westmeier2022}. ASKAP has a 30-square-degree instantaneous field of view and, for WALLABY, provides a spatial and spectral resolution of 30\,arcsec and 4\,km\,s$^{-1}$, respectively, and can reach a nominal sensitivity of 1.6\,mJy per beam per 4\,km\,s$^{-1}$ channel with a 16\,h integration. The pilot survey phase 1 of WALLABY has observed six 30-square-degree tiles. This includes two pairs of adjacent tiles, each covering 60-square-degrees, in the directions of the Hydra~I cluster and the NGC\,4636 galaxy group \citep{Westmeier2022}.

In this work, we investigate and quantify variations in the H\,\textsc{i} radial surface density profiles of $\sim280$ nearby, gas-rich galaxies by the average H\,\textsc{i} surface density and size of the H\,\textsc{i} disc and look at how the structure of the stellar disc (quantified by stellar mass and stellar mass surface density) and star formation affect these quantities. In Section~\ref{sec:data}, we describe the data and derive the physical quantities we use for this work. We describe and characterise our sample in Section~\ref{sec:sample_selection} and present our results in Section~\ref{sec:results}. We place our results in context with previous work and discuss possible physical drivers in Section~\ref{sec:discussion}. We summarise our conclusions in Section~\ref{sec:conclusion}. Throughout, we adopt optical velocities (c$z$) in the heliocentric reference frame, the AB magnitude convention and we assume a flat $\Lambda$CDM cosmology with $H_0=67.7$\,km\,s$^{-1}$\,Mpc$^{-1}$ \citep{Planck2016}.

\section{Data}
\label{sec:data}

\subsection{WALLABY}
\label{s-sec:hi_data}

\begin{table*}
	\centering
	\caption{Sky coverage and number of galaxies with resolved isodensity ($R_{\rm{iso,HI}}$, ISO sample) and effective ($R_{\rm{50,HI}}$, EFF sample) radii in the Hydra and NGC\,4636 fields.}
	\label{table:sample_param}
	\begin{tabular}{lcccccr}
	    \hline
		            & RA (J2000) & Dec (J2000) & \multicolumn{2}{c}{$R>30$\,arcsec} & \multicolumn{2}{c}{$R>20$\,arcsec} \\
		            & [deg] & [deg] & $R_{\rm{iso,HI}}$  & $R_{\rm{50,HI}}$ & $R_{\rm{iso,HI}}$  & $R_{\rm{50,HI}}$   \\
		\hline
		Hydra       & [$150.75$, $163.25$] & [$-30.5$, $-24.5$]  & 181  & 42  & 190  & 100  \\
		NGC\,4636   & [$186.75$, $193.00$] & [$-4$, $+8$]        & 85   & 41  & 91   & 64   \\
		Total       &                      &                     & 266  & 83  & \textbf{281}  & \textbf{164} \\
		Sample      &                      &                     &      &     & ISO  & EFF \\
		\hline
	\end{tabular}
\end{table*}

In this work, we use data from the first WALLABY pilot survey Public Data Release \citep[PDR1,][]{Westmeier2022} covering two 60-square-degree fields in the directions of the Hydra cluster and the NGC\,4636 group. For details on the WALLABY observations and data processing see \cite{Westmeier2022}. Source catalogues of H\,\textsc{i} detections were produced for each field using the Source Finding Application 2 \citep[SoFiA2,][]{Westmeier2021} and released internally to the WALLABY team. We use H\,\textsc{i} detections from the WALLABY PDR1 denoted Hydra TR2 and NGC\,4636 TR1 (we list the sky coverage in Table~\ref{table:sample_param}). Hydra TR2 was produced from blind source finding using a median-based approximation of the root mean square (RMS) detection threshold of $3.5\sigma_{\mathrm{RMS}}$ (which is more robust against outliers than the RMS), where $\sigma_{\mathrm{RMS}}$ is the RMS level of the H\,\textsc{i} spectral line cube (nominally $\sigma_{\mathrm{RMS}}=1.6$\,mJy\,beam$^{-1}$, but it varies spatially and spectrally, see \citeauthor{Westmeier2022} \citeyear{Westmeier2022}). For NGC\,4636 TR1, the source finding used optical and H\,\textsc{i} position and redshift priors and a higher detection threshold of $4\sigma_{\mathrm{RMS}}$ due to artefacts and bright continuum residuals in the H\,\textsc{i} spectral line cube, which made blind source finding impossible (see \citeauthor{Westmeier2022} \citeyear{Westmeier2022} for full details on the source finding run parameters and the WALLABY PDR1 source catalogue\footnote{The WALLABY PDR1 source catalogue and associated source data products (e.g.\ integrated spectra and moment maps) are available through the CASDA ASKAP Science Data Archive \citep[CASDA,][]{Chapman2015, Huynh2020} using the DOI \url{https://doi.org/10.25919/09yg-d529}.}).

The final source catalogue contains a total of 414 detections of H\,\textsc{i} emission in the two fields with integrated signal-to-noise ratios $\mathrm{SNR}>4$. We note that not all these detections correspond to individual, real sources as some artefacts remain in the catalogue, SoFiA has detected multiple H\,\textsc{i} components for single sources (i.e.\ equivalent to shredding in optical imaging) and some detections contain multiple galaxies within a single H\,\textsc{i} envelope (i.e.\ close interacting systems). We limit our sample to only those H\,\textsc{i} detections with single optical counterparts with multi-wavelength (ultraviolet, optical and near-infrared) coverage ($N=282$, see Section~\ref{sec:sample_selection}).

We use the method described in \cite{Reynolds2022} to calculate the H\,\textsc{i} mass and measure the H\,\textsc{i} radius. In summary, we fit a 2-dimensional Gaussian to the SoFiA produced moment 0 map, which we use to define annuli within which we measure the H\,\textsc{i} surface density and use to produce the radial H\,\textsc{i} surface density profile. However, we make the following changes to the published method. The majority of WALLABY sources are detected due to spatially and spectrally smoothing the spectral line cube at various scales with SoFiA to increase the cube SNR. As a result, these sources may not be deconvolved as they are too faint to be detected in individual channels during the deconvolution step of the ASKAPsoft processing pipeline \citep[for further details see][]{Westmeier2022}. This causes the majority of integrated H\,\textsc{i} flux measurements to be underestimated, with fainter H\,\textsc{i} detections being more severely affected. We apply the statistical correction to the integrated fluxes proposed in \cite{Westmeier2022} presented in their equations~9 and 10. The correction is $<10$\% for fluxes $>10^5$\,Jy\,Hz, but increases from $\sim25$\% to $\sim60$\% as the measured flux decreases from $\sim10^{4.5}$ to $\sim10^3$\,Jy\,Hz. We also correct the radial surface brightness profiles measured from the moment 0 maps by scaling the profiles by the ratio between the measured and corrected integrated fluxes. 

We then convert the profile from surface brightness to surface density \citep[e.g.][]{Meyer2017} and measure the isodensity radius at $1\,\rm{M}_{\odot}\,\rm{pc}^{-2}$ ($R_{\rm{iso,HI}}$), and the half-mass (effective, $R_{\rm{50,HI}}$) radius. We deconvolve these measured radii by the ASKAP synthesised beam of $\sim30$\,arcsec to get the final $R_{\rm{iso,HI}}$ and $R_{\rm{50,HI}}$ values (where we assume that the beam and H\,\textsc{i} disc can be approximated as Gaussians). We calculate the average H\,\textsc{i} surface density, $\mu_{\mathrm{HI}}$, (hereafter referred to as the H\,\textsc{i} surface density) within the both the isodensity ($\mu_{\mathrm{iso,HI}}$) and effective ($\mu_{\mathrm{50,HI}}$) radii by integrating the H\,\textsc{i} flux in the moment 0 maps contained within elliptical apertures with the semi-major axis defined by the H\,\textsc{i} radius and dividing by the area of the aperture. We then convert to units of surface density ($\mathrm{M}_{\odot}\,\mathrm{pc}^{-2}$). We correct all surface density measurements that we describe above for inclination (i.e.\ deproject to face-on). We determine the inclination from the ratio of the minor to major axes of the 2-dimensional Gaussian fit to the H\,\textsc{i} moment 0 map that we use to extract the radial surface density profile (described in \citeauthor{Reynolds2022} \citeyear{Reynolds2022}).

Our calculated H\,\textsc{i} surface densities assume that the H\,\textsc{i} gas is optically thin. If the gas is not optically thin, then the H\,\textsc{i} surface densities will be underestimated when we correct for inclination. However, we are also likely under-estimating the inclination for poorly resolved galaxies (i.e.\ $<4$ beams) due to the size of the minor axis being comparable to that of the synthesised beam. This introduces a bias towards lower inclinations for poorly resolved galaxies, which reduces the effect from correcting for inclination and counters the underestimate introduced by the assumption that the gas is optically thin. We find that the measured inclination does not correlate with either $\mu_{\mathrm{iso,HI}}$ or $\mu_{\mathrm{50,HI}}$. We conclude that the inclination correction does not introduce a systematic bias in the H\,\textsc{i} surface densities. However, it likely contributes to the scatter in the measured H\,\textsc{i} surface densities.

\subsection{Ancillary Data}
\label{s-sec:ancillary}

We follow the method described in \cite{Reynolds2022} to derive stellar masses ($M_*$) and optical sizes from PanSTARRS \citep{Chambers2016,Flewelling2016} $g$- and $r$-bands images and total star formation rates (SFRs) from \textit{GALEX} \citep{Martin2005,Morrissey2007} near-UV (NUV) and WISE \citep{Wright2010} W3-/W4-band magnitudes. However, we have made the following changes to the measured PanSTARRS photometry concerning local sky background removal.

Initially, we measure the photometry for the PanSTARRS $r$-band image as described in \cite{Reynolds2022} in which the centre, position angle and inclination angle of each isophotal annulus are allowed to vary. We then fix the annulus centre, inclination angle and position angle to the values for the 25\,mag\,arcsec$^{-2}$ isophote from the initial measurement. This also provides an initial estimate of the $r$-band radius, $R_{\mathrm{r,guess}}$. We measure the background at $>1.5R_{\mathrm{r,guess}}$ in five annuli of width $0.1R_{\mathrm{r,guess}}$ and take the average of these annuli as the mean background. We subtract the mean background and remeasure the photometry using annuli with the centre, inclination angle and position angle fixed. We define the isophotal radius using the 25\,mag\,arcsec$^{-2}$ isophote ($R_{\rm{iso,r}}$, i.e.\ the isophote we use to define the aperture for measuring the total magnitude in the $g$- and $r$-bands). The local background subtraction has only a minor effect on the measured total magnitudes ($\sim0.01\pm0.07$\,mag brighter). We measure the effective radius ($R_{\rm{50,r}}$) at the point containing half the flux within the aperture defined by the 26\,mag\,arcsec$^{-2}$ isophote (i.e.\ 1 magnitude fainter than the surface brightness we use for the isophotal radius). We also measure the stellar mass surface density ($\mu_*$) using the derived stellar mass and effective $r$-band radius as
\begin{equation}
    \mu_*=M_*/(2\pi R_{50,r}^2),
    \label{equ:star_surface_density}
\end{equation}
where $R_{\rm{50,r}}$ is in units of kpc.

\begin{figure*}[ht]
	\includegraphics[width=\textwidth]{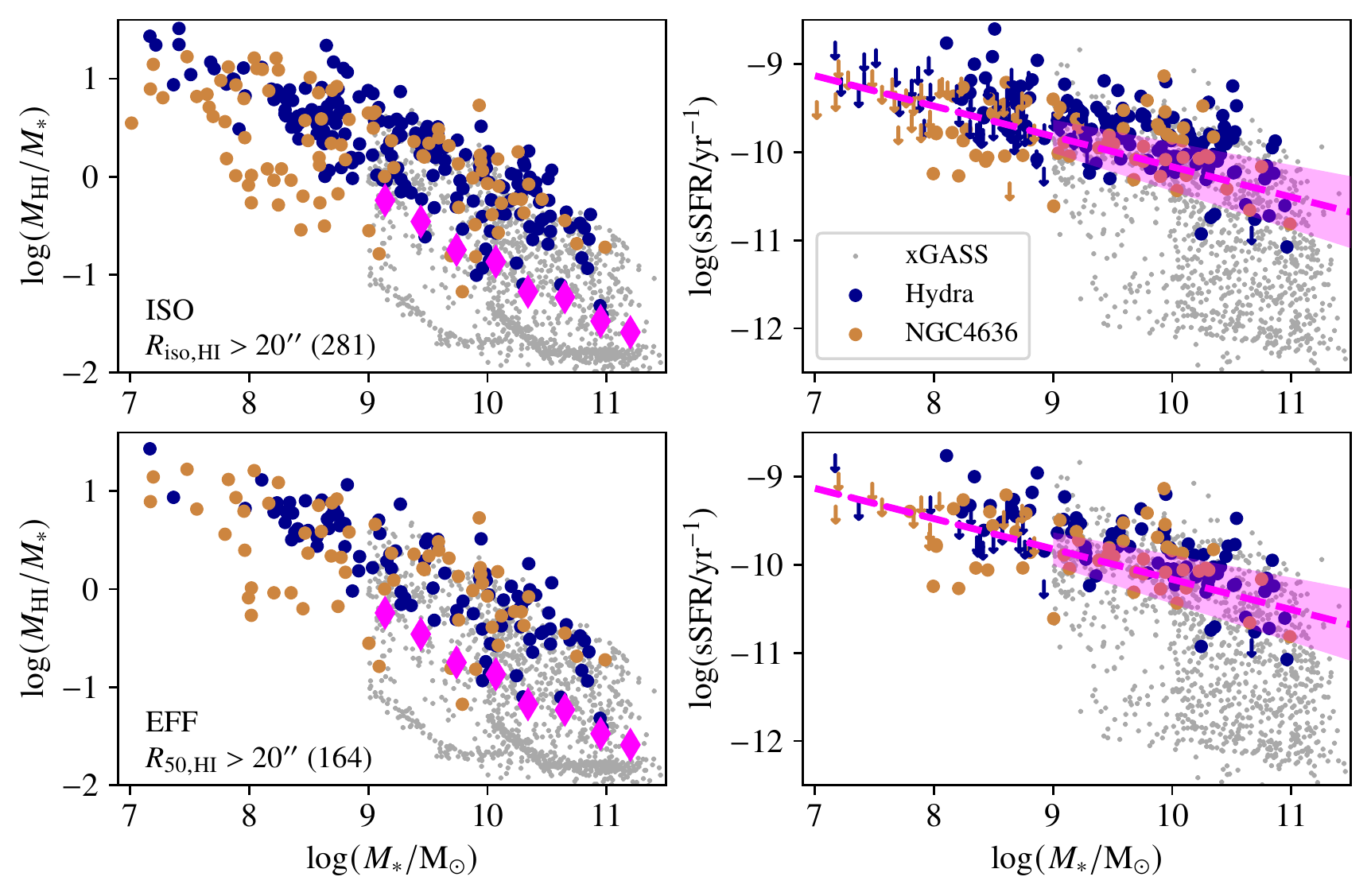}
    \caption{H\,\textsc{i} gas fraction ($M_{\mathrm{HI}}/M_*$, left column) and specific star formation rate (sSFR, right column) vs stellar mass ($M_*$) for galaxies with $R_{\rm{iso,HI}}>20$\,arcsec and with $R_{\rm{50,HI}}>20$\,arcsec (top and bottom rows, respectively). The Hydra and NGC\,4636 field galaxies are indicated in blue and orange, respectively, and the grey points are the xGASS sample. The magenta diamonds in the left column are the median gas fractions for xGASS from \citet{Catinella2018}. In the left column, the numbers in the upper right corner indicate the number of galaxies detected in the Hydra and NGC\,4636 fields and the number in the lower left corner gives the total number of galaxies in the resolved sample. The magenta dashed line and shaded region in the right column are the star forming main sequence (SFMS) and $1\sigma$ scatter in the SFMS from \citet{Janowiecki2020}. In the right column, the down pointing arrows indicate galaxies with sSFR upper limits. }
    \label{fig:sample_selection}
\end{figure*}

\section{Sample Selection and Characterisation}
\label{sec:sample_selection}

Our goal is to measure the H\,\textsc{i} disc sizes and average surface densities in WALLABY galaxies and how these properties relate to stellar structure (which we characterise by the galaxies' stellar mass and stellar mass surface density) and star formation. To achieve this, we require the H\,\textsc{i} detections to be good quality and spatially resolved (i.e.\ no artefacts in the sources) with single optical counterparts (i.e.\ no closely interacting systems of multiple optical counterparts contained within a single H\,\textsc{i} detection) and have multi-wavelength coverage in PanSTARRS $g$- and $r$-bands, \textit{GALEX} NUV-band and WISE W1 and W3/W4-bands. Of the original 414 WALLABY detections, 56 contain artefacts or are closely interacting systems contained within a single H\,\textsc{i} envelope. A further 76 either have incomplete PanSTARRS coverage, are impacted by foreground stars, are not detected in the PanSTARRS images or are missing \textit{GALEX} coverage. 

These cuts result in a sample of 282 galaxies, of which 281 are at least marginally resolved with deconvolved isodensity H\,\textsc{i} radii $R_{\mathrm{iso,HI}}>20$\,arcsec (i.e.\ $>1.3$\,beams along the major axis). The sample decreases to 164 galaxies with effective H\,\textsc{i} radii $R_{\mathrm{50,HI}}>20$\,arcsec. If we further restrict our sample to slightly better spatially resolved galaxies with radii $>30$\,arcsec (i.e.\ $>2$\,beams along the major axis), the sample reduces to 266 (83) with resolved isodensity (effective) radii. We present results for the galaxy samples with $R_{\mathrm{iso,HI}}>20$\,arcsec ($N=281$ galaxies, hereafter referred to as the ISO sample) and $R_{\mathrm{50,HI}}>20$\,arcsec ($N=164$ galaxies, hereafter referred to as the EFF sample) as these provide the best statistics, most notably for galaxies with resolved effective radii (i.e.\ nearly double the sample size, see Table~\ref{table:sample_param}). We find similar results if we use the radii $>30$\,arcsec sample.

As a blind H\,\textsc{i} survey, the galaxy population that WALLABY detects is inherently biased (i.e.\ with increasing distance only larger and more gas-rich galaxies are easily detected). To put the WALLABY detections into a broader context, we compare the H\,\textsc{i} gas fraction ($M_{\mathrm{HI}}/M_*$) and specific star formation rate (sSFR) of our sample with the xGASS \citep{Catinella2018} representative sample (left and right columns of Figure~\ref{fig:sample_selection}, respectively). Unlike blind H\,\textsc{i} surveys, xGASS is not biased towards H\,\textsc{i} rich systems as it is a stellar mass selected ($10^9<M_*<10^{11}$\,M$_{\odot}$), gas fraction limited ($M_{\mathrm{HI}}/M_*>0.02$--0.1) survey. We show the ISO and EFF samples in the top and bottom rows, respectively. We distinguish between galaxies detected in the Hydra and NGC\,4636 fields (blue and orange, respectively) as different source finding techniques were applied to each field (Section~\ref{s-sec:hi_data} and \citeauthor{Westmeier2022} \citeyear{Westmeier2022}) and the detected galaxies in each field are concentrated at different distances ($\sim55$\,Mpc vs $\sim20$\,Mpc, respectively).

At fixed stellar mass, the galaxies detected by WALLABY predominantly have H\,\textsc{i} gas fractions above the xGASS medians (\citeauthor{Catinella2018} \citeyear{Catinella2018}, magenta diamonds in the left panels of Figure~\ref{fig:sample_selection}). This illustrates that the majority of the WALLABY detections are gas-rich. \cite{Janowiecki2020} defined a star forming main sequence (SFMS) and its $1\sigma$ scatter using xGASS, which we overlay in the right panels of Figure~\ref{fig:sample_selection} (magenta dashed line and shaded region). For illustrative purposes, we extend the SFMS down to $M_*=10^{7}\,\mathrm{M}_{\odot}$ (i.e.\ below the xGASS stellar mass limit of $M_*=10^{9}\,\mathrm{M}_{\odot}$), but only show the scatter for the xGASS stellar mass range. WALLABY detects mostly star forming galaxies that predominantly lie either above or within the scatter of the SFMS, with only a small population of galaxies with sSFR below the $1\sigma$ scatter in the SFMS. For 83 galaxies in the ISO sample, the derived sSFRs are classified as upper limits if they are not detected (i.e.\ $\mathrm{SNR}<5$) in either the \textit{GALEX} NUV-band or both the WISE W3- and W4-band images (indicated by the down pointing arrows in the right panels of Figure~\ref{fig:sample_selection}).

There is a population of galaxies from the NGC\,4636 field below $M_*\lesssim10^{10}\,\mathrm{M}_{\odot}$ that are offset to lower gas fractions ($\sim0.5$\,dex lower), but no corresponding population of galaxies in the Hydra field. Above $M_*\gtrsim10^{10}\,\mathrm{M}_{\odot}$, there are $\sim6$ galaxies from the Hydra field with similarly offset gas fractions. This is due to the NGC\,4636 field containing more nearby galaxies (i.e.\ $<30$\,Mpc) compared to the Hydra field (i.e.\ $>35$\,Mpc) and the targeted source finding strategy applied to the NGC\,4636 field (Section~\ref{s-sec:hi_data}). Compared to the ISO sample, the EFF sample is slightly biased towards galaxies with $M_*>10^{9}\,\mathrm{M}_{\odot}$ (i.e.\ for $M_*<10^{9}\,\mathrm{M}_{\odot}$ and $M_*>10^{9}\,\mathrm{M}_{\odot}$, $\sim50\%$ vs $\sim65\%$ of galaxies in our sample have $R_{\mathrm{50,HI}}>20$\,arcsec, respectively).

\section{Results}
\label{sec:results}

As we discuss in the introduction, many previous studies targeted relatively small numbers of galaxies selected to have high spatial resolution and were able to probe radial variations in the H\,\textsc{i} distribution in the galaxies' inner regions (i.e.\ within the optical disc) with other galaxy properties on a galaxy-by-galaxy basis. Unlike these data sets, our sample is limited to the galaxies that we detect in the observed field, which includes a small fraction of well-resolved galaxies (i.e.\ $>4$ beams along the major axis; $\sim25\%$ and $\sim10\%$ of galaxies' isodensity and effective sizes are resolved by $>6$ beams, respectively) and a large fraction of marginally resolved galaxies (i.e.\ $<4$ beams along the major axis; $\sim75\%$ and $\sim90\%$ of galaxies for the ISO and EFF samples, respectively). Due to the poor resolution of the majority of our sample, we are unable to probe radial variations in the H\,\textsc{i} on small scales. Instead, we parameterise the H\,\textsc{i} spatial distribution using quantities that we can measure for marginally resolved galaxies: the size of the H\,\textsc{i} disc and the H\,\textsc{i} surface density within the H\,\textsc{i} disc. 

We also investigate the effect of spatial resolution on the measured radii and surface densities by convolving the moment 0 maps of galaxies resolved by $>2$ beams to an equivalent resolution of 2 beams along the major axis. We find a difference smaller than 10\% between the structural parameters measured from the two sets of maps (see \ref{appendix:beam_resolution}), confirming that the findings presented below are not significantly affected by differences in resolution within our sample.

\begin{figure*}[ht]
	\includegraphics[width=\textwidth]{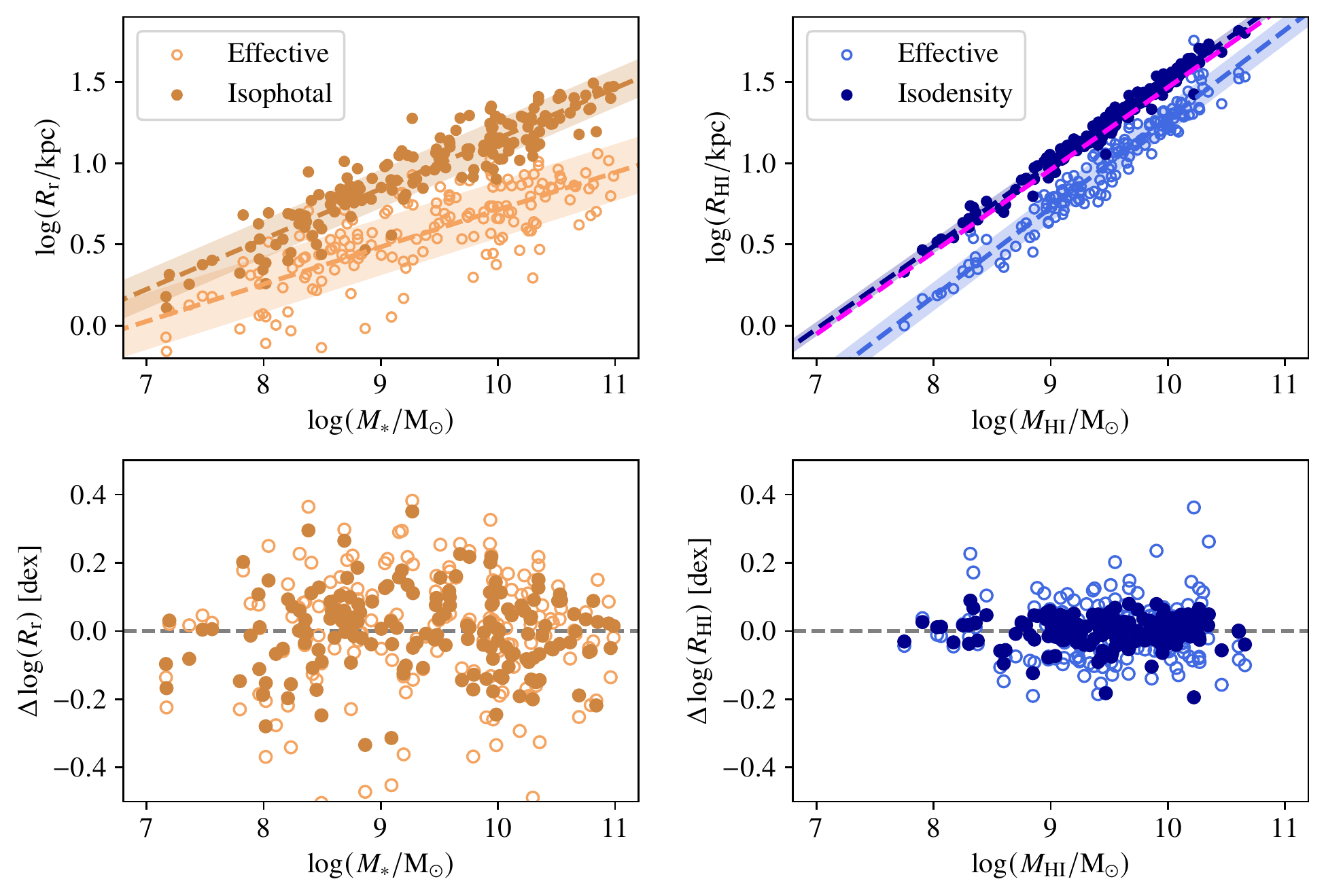}
    \caption{Top: Size-mass relations for $r$-band/stellar (left) and H\,\textsc{i} (right) discs. The isophotal/isodensity radii are plotted as filled, dark circles and the effective radii as unfilled, light circles. The dashed lines and shaded regions show the linear least-square regression best fit relations and the $1\sigma$ scatter for each radius definition. The dashed magenta line in the top right panel is the best fit H\,\textsc{i} size-mass relation from \citet{Wang2016}. Bottom: Residual vertical offsets of galaxies from the best fit size-mass relations. }
    \label{fig:size_mass_residual}
\end{figure*}

\subsection{Size-Mass Relations}
\label{s-sec:size_mass_relations}

Studies show that galaxies' radial H\,\textsc{i} surface density profiles normalised by the H\,\textsc{i} radius defined at $1\,\rm{M}_{\odot}\,\rm{pc}^{-2}$, $R_{\rm{iso,HI}}$, are approximately exponential in the outer disc (i.e.\ $r\gtrsim0.5R_{\rm{iso,HI}}$, e.g.\ \citeauthor{Swaters2002} \citeyear{Swaters2002}, \citeauthor{Wang2016} \citeyear{Wang2016}). At smaller radii (e.g.\ $\lesssim0.5R_{\rm{iso,HI}}$), the radial H\,\textsc{i} surface density is found to vary by $\sim1$\,dex from $\sim1$--$10\,\rm{M}_{\odot}\,\rm{pc}^{-2}$. In comparison with the H\,\textsc{i}, the stellar mass surface density shows a larger variation (e.g.\ $\sim2$\,dex, \citeauthor{Catinella2018} \citeyear{Catinella2018}), which is due to the presence or lack of a central brightness concentration or bulge-like component. Variations in the central surface brightness and internal extinction contribute to the difference in slope and scatter in stellar size-mass relations derived using isophotal vs effective radii \citep[e.g.][]{Saintonge2011, Cortese2012, MunozMateos2015, Trujillo2020, Sanchez-Almeida2020}. Although galaxies do not have similar central concentrations of H\,\textsc{i}, variations in the inner H\,\textsc{i} surface density may produce differences between H\,\textsc{i} size-mass relations derived from isodensity vs effective radii.

We plot the optical $r$-band and H\,\textsc{i} size-mass relations for our EFF sample in Figure~\ref{fig:size_mass_residual} (left and right panels of top row, respectively). We plot both effective (light, unfilled circles) and isophotal/isodensity (dark, filled circles) defined radii and fit linear least-squares regression to the size-mass relations. The best fit parameters and $1\sigma$ scatter (dashed lines and shaded regions) are listed in Table~\ref{table:size_mass_coef}. The lower panels show the offsets from the corresponding size-mass relations in the top row. The stellar and isodensity H\,\textsc{i} relations are consistent within uncertainties if we use the ISO sample.

Focusing first on the stellar size-mass relation (top left panel), we find that isophotal sizes produce a steeper and less scattered relation than effective sizes ($a=0.31$ vs 0.23 and $\sigma=0.12$ vs 0.17, respectively). The bottom left panel shows that individual galaxies scatter significantly around both relations (i.e.\ $\sim0.4$\,dex and $\sim0.5$\,dex for isophotal and effective radii, respectively). The differing slopes and scatters we measure for the isophotal and effective stellar size-mass relations are consistent with the literature \citep[e.g.][]{MunozMateos2015, Cortese2012, Trujillo2020, Sanchez-Almeida2020}.

Our results for the stellar relations indicate that effective radii increase more slowly than isophotal radii with increasing stellar mass and that the isophotal size is a better tracer for a galaxy's total stellar mass. Both of these effects are a result of the sensitivity of effective radii to central stellar concentrations, which are more prominent in higher mass galaxies. In our galaxy sample, the central surface brightnesses of the $r$-band radial profiles span $\sim19$--25\,mag\,arcsec$^{-2}$ (a factor of $\sim250$). We note that our sample contains predominantly late-type galaxies, few of which are likely to contain a true bulge component (i.e.\ using $R_{\mathrm{90,r}}/R_{\mathrm{50,r}}$ as a proxy for the bulge to total ratio, $\sim85\%$ of our sample has $R_{\mathrm{90,r}}/R_{\mathrm{50,r}}<2.5$ which is consistent with being disc-dominated systems, e.g.\ \citeauthor{Ferreras2005} \citeyear{Ferreras2005}). The majority of galaxies with $R_{\mathrm{90,r}}/R_{\mathrm{50,r}}>2.5$ have $M_*>10^{9}\,\mathrm{M}_{\odot}$.

Moving to the H\,\textsc{i} size-mass relations (top right panel of Figure~\ref{fig:size_mass_residual}), we find that isodensity sizes produce a shallower and less scattered relation than effective sizes ($a=0.51$ vs 0.55 and $\sigma=0.04$ vs 0.09, respectively). The bottom right panel highlights the tightness of both H\,\textsc{i} relations with few galaxies scattering by more than 0.1--0.2\,dex. Our isodensity H\,\textsc{i} size-mass relation agrees well with the \cite{Wang2016} relation (magenta line in the top right panel; $a_{\mathrm{W16}}=0.506$, $b_{\mathrm{W16}}=3.594$).\footnote{Note that the \cite{Wang2016} relation uses the H\,\textsc{i} diameter and we have scaled the $b_{\mathrm{W16}}$ value to account for our use of the H\,\textsc{i} radius (i.e.\ subtracting $\log(2)$).} However, our relation is tighter ($\sigma=0.04$ vs 0.06), which may be due to the uniform WALLABY sample compared to the compilation of interferometric H\,\textsc{i} surveys used by \cite{Wang2016}. A similar increase in scatter for the effective vs isodensity H\,\textsc{i} size-mass relation is also found in simulations \citep{Stevens2019b}.

The steeper effective H\,\textsc{i} size-mass relation indicates that, unlike the stellar equivalent, H\,\textsc{i} effective radii grow faster than isodensity radii with increasing mass. The difference in slopes is small, but significant (i.e.\ $\sim4$ times the uncertainty in the measured slopes, Table~\ref{table:size_mass_coef}). Similar to the results for the stellar component, isodensity H\,\textsc{i} radii better trace the total H\,\textsc{i} mass than effective radii. However, effective H\,\textsc{i} radius is still a fairly good tracer as the scatter is less than that for either stellar size-mass relations. The smaller differences between the H\,\textsc{i} size-mass relations compared to the stellar relations discussed above are likely due to the the exponential shape of the outer H\,\textsc{i} profiles and the lower dynamic range of the central H\,\textsc{i} surface densities (i.e.\ spanning $\sim1$\,dex).

We conclude that the increased scatter in the effective H\,\textsc{i} size-mass relation is due to variations in the H\,\textsc{i} spatial distribution in the central regions (i.e.\ within the effective radius) of galaxies (e.g.\ illustrated for early- vs late-type galaxies by \citeauthor{Wang2016} \citeyear{Wang2016} in their figure~2). 

In the following sections, we investigate variations in the size of the H\,\textsc{i} disc normalised to the optical $r$-band disc size, $R_{\mathrm{HI}}/R_{\mathrm{r}}$, and the H\,\textsc{i} surface density, $\mu_{\mathrm{HI}}$, for both the isodensity and effective radii. We investigate how these quantities correlate with stellar mass, stellar mass surface density, sSFR, $\mathrm{NUV}-r$ colour and the offset from the star forming main sequence ($\Delta\,\rm{SFMS}$) defined using xGASS by \cite{Janowiecki2020}, shown in the right panels of Figure~\ref{fig:sample_selection} by the dashed magenta line, to identify possible drivers responsible for producing differences in the H\,\textsc{i} spatial distributions.

\begin{table}
	\centering
	\caption{Parameters of the linear least-squares regression best fit size-mass relations of the form $\log(y/\mathrm{kpc}) = a \log(x/\mathrm{M}_{\odot}) - b$. The scatter in the relations and Pearson correlation coefficients are given by $\sigma$ and $\rho$, respectively.}
	\label{table:size_mass_coef}
	\begin{tabular}{llcccr}
	    \hline
		$x$             & $y$                 & $a$            & $b$              & $\sigma$  & $\rho$ \\
		\hline
		$M_*$           & $R_{\rm{iso,r}}$    & $0.31\pm0.01$  & $1.94\pm0.09$   & $0.12$    & $0.92$ \\
		                & $R_{\rm{50,r}}$     & $0.23\pm0.02$  & $1.58\pm0.14$   & $0.17$    & $0.78$ \\
    	$M_{\rm{HI}}$   & $R_{\rm{iso,HI}}$   & $0.51\pm0.01$  & $3.58\pm0.05$   & $0.04$    & $0.99$ \\
		                & $R_{\rm{50,HI}}$    & $0.55\pm0.01$  & $4.19\pm0.11$   & $0.09$    & $0.97$ \\
		\hline
	\end{tabular}
\end{table}

\begin{figure*}
	\includegraphics[width=\textwidth]{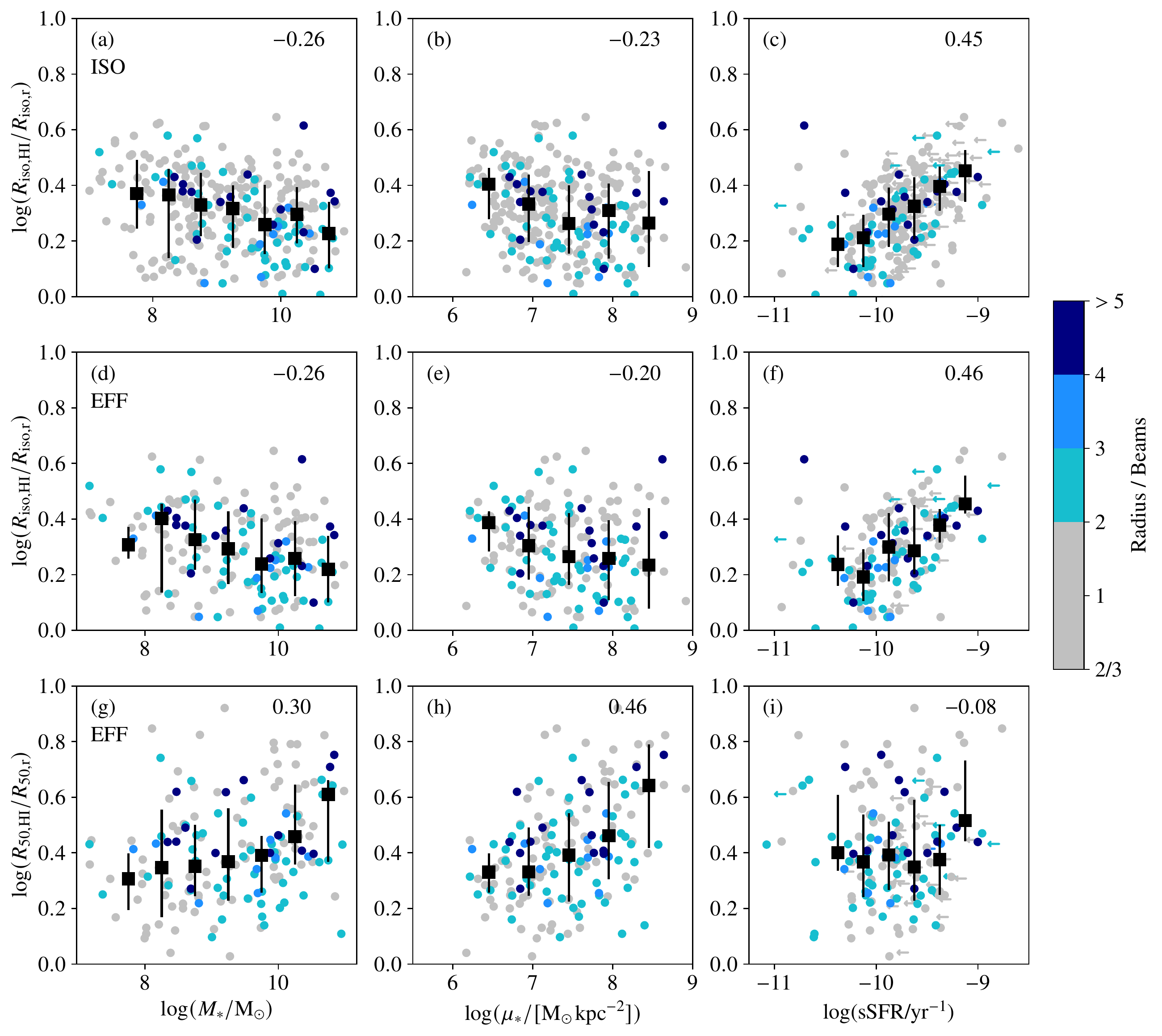}
    \caption{The H\,\textsc{i} radius normalised by the $r$-band radius ($R_{\rm{HI}}/R_{\rm{r}}$) plotted against stellar mass ($M_*$), stellar mass surface density ($\mu_*$) and specific star formation rate (sSFR; columns from left to right, respectively). The top row shows the normalised isodensity H\,\textsc{i} disc size ($R_{\rm{iso,HI}}/R_{\rm{iso,r}}$) for the ISO sample. The middle row shows $R_{\rm{iso,HI}}/R_{\rm{iso,r}}$ for the EFF sample. The bottom row shows the normalised effective H\,\textsc{i} disc size ($R_{\rm{50,HI}}/R_{\rm{50,r}}$) for the EFF sample. The corresponding Pearson correlation coefficients (top right corner of each panel) are tabulated in the first three rows of Table~\ref{table:correlations}. The circles are coloured by the radius expressed in terms of the number of beams (e.g.\ $R_{\mathrm{iso,HI}}=60$\,arcsec is 2 beams). The black squares show binned medians with the error bars showing the 20$^{\rm{th}}$ and 80$^{\rm{th}}$ percentiles (all bins contain $>5$ galaxies). In the right column (sSFR), we indicate galaxies whose sSFRs are upper limits by left pointing arrows.}
    \label{fig:all_sratio}
\end{figure*}

\subsection{The Normalised HI Disc Size}
\label{s-sec:hi_size}

Beyond the effective radius H\,\textsc{i} radial profiles are approximately exponential. We probe the outer part of the radial H\,\textsc{i} profile using the size of the H\,\textsc{i} disc relative to the optical disc, which provides a measure of the size of the H\,\textsc{i} gas reservoir. H\,\textsc{i} beyond the optical disc will not directly impact the inner star-forming disc, but provides a reservoir that has the potential to be funnelled towards the galaxy centre for future star formation \citep[e.g.][]{Schmidt2016}. We normalise the H\,\textsc{i} radius by the equivalent $r$-band radius (i.e.\ isodensity H\,\textsc{i} by isophotal $r$-band, $R_{\rm{iso,HI}}/R_{\rm{iso,r}}$, and effective H\,\textsc{i} by effective $r$-band, $R_{\rm{50,HI}}/R_{\rm{50,r}}$). 

In Figure~\ref{fig:all_sratio}, we show how $R_{\rm{iso,HI}}/R_{\rm{iso,r}}$ varies as a function of $M_*$, $\mu_*$ and sSFR (top row, panels a--c, respectively) for the ISO sample. To guide the reader's eye, we plot medians with error bars indicating the 20$^{\rm{th}}$ and 80$^{\rm{th}}$ percentiles as black squares (all bins contain $>5$ galaxies). We quantify the level of correlation between each pair of quantities using the Pearson correlation coefficient ($\rho$), which we show in the upper right corner of each panel and list in the first row of Table~\ref{table:correlations}. The significance of the correlations is indicated by the $p$-values (in brackets), where the correlation is deemed significant if the $p$-value is $<0.05$. We note that the Pearson correlations are calculated on the data directly, not from the medians. Although we do not plot $R_{\rm{iso,HI}}/R_{\rm{iso,r}}$ as a function of $\mathrm{NUV}-r$ colour and $\Delta\,\rm{SFMS}$, we include the Pearson correlations in Table~\ref{table:correlations}. We indicate galaxies with upper limit derived SFRs by the left pointing arrows in panel c. A large fraction of our sample is marginally resolved, which may bias our results by over-estimating the H\,\textsc{i} sizes. We indicate spatially resolved galaxies with radii $>2$ beams in progressively darker shades of blue (galaxies with radii $<2$ beams are in grey). There is no apparent correlation between the derived galaxy quantities and either poorly or well resolved galaxies and we conclude that spatial resolution is not introducing artificial correlations. 

\begin{table*}
	\centering
	\caption{Pearson correlation coefficients, $\rho$, and $p$-values (in brackets) for the isodensity and effective $R_{\rm{HI}}/R_{\rm{r}}$ and $\mu_{\rm{HI}}$ as functions of $M_*$, $\mu_*$, sSFR, $\mathrm{NUV}-r$ colour and $\Delta\,\rm{SFMS}$ (See Figures~\ref{fig:all_sratio} and \ref{fig:all_muhi}). The number of galaxies used to measure the Pearson correlations is given by $N$.}
	\label{table:correlations}
	\begin{tabular}{lccccccr}
	    \hline
		                                  & Sample & $N$   & $M_*$            & $\mu_*$           & sSFR            & $\mathrm{NUV}-r$   & $\Delta\,\rm{SFMS}$ \\
		\hline
		$R_{\rm{iso,HI}}/R_{\rm{iso,r}}$ & ISO & $281$   & $-0.26$ ($<0.01$) & $-0.23$ ($<0.01$) & $0.45$ ($<0.01$)  & $-0.45$ ($<0.01$)  & $0.30$ ($<0.01$)  \\
		$R_{\rm{iso,HI}}/R_{\rm{iso,r}}$ & EFF & $164$   & $-0.26$ ($<0.01$) & $-0.20$ ($<0.01$) & $0.46$ ($<0.01$)  & $-0.50$ ($<0.01$)  & $0.36$ ($<0.01$)  \\
		$R_{\rm{50,HI}}/R_{\rm{50,r}}$ & EFF   & $164$   & $0.20$  ($<0.01$) & $0.39$  ($<0.01$) & $0.10$ ($0.26$)   & $0.06$ ($0.42$)    & $0.20$ ($0.01$)  \\
		\\
		$\mu_{\rm{iso,HI}}$              & ISO & $281$   & $-0.24$ ($<0.01$) & $-0.20$ ($<0.01$) & $0.33$ ($<0.01$)  & $-0.31$ ($<0.01$)  & $0.18$ ($<0.01$)  \\
		$\mu_{\rm{iso,HI}}$             & EFF  & $164$   & $-0.32$ ($<0.01$) & $-0.28$ ($<0.01$) & $0.48$ ($<0.01$)  & $-0.44$ ($<0.01$)  & $0.30$ ($<0.01$) \\
		$\mu_{\rm{50,HI}}$               & EFF & $164$   & $-0.36$ ($<0.01$) & $-0.29$ ($<0.01$) & $0.50$ ($<0.01$)  & $-0.44$ ($<0.01$)  & $0.30$ ($<0.01$)  \\ 
		\hline
	\end{tabular}
\end{table*}

The normalised isodensity H\,\textsc{i} radius correlates weakly with the stellar properties: $M_*$ and $\mu_*$ ($\rho=-0.26$ and $-0.23$, panels a and b). Galaxies with higher stellar masses and/or higher stellar surface densities tend to have marginally less extended H\,\textsc{i} discs for their optical sizes. The correlations are stronger between $R_{\rm{iso,HI}}/R_{\rm{iso,r}}$ and sSFR ($\rho=0.45$, panel c). The $\mathrm{NUV}-r$ colour and $\Delta\,\rm{SFMS}$ are also show stronger correlations with $R_{\rm{iso,HI}}/R_{\rm{iso,r}}$ (see Table~\ref{table:correlations}). These correlations indicate that galaxies with more extended H\,\textsc{i} reservoirs (relative to their stellar discs) tend to be more star forming. The weaker correlation with $\Delta\,\rm{SFMS}$ compared to sSFR indicates that the correlation between $R_{\rm{iso,HI}}/R_{\rm{iso,r}}$ and $M_*$ contributes to the strong correlation with sSFR as sSFR has a weak dependence on stellar mass (e.g.\ Figure~\ref{fig:sample_selection}). However, the weakened correlation with $\Delta\,\rm{SFMS}$ may also be due to the bias of our galaxy sample toward star forming galaxies on or above the SFMS (i.e.\ we have poor statistics below the SFMS, see Section~\ref{sec:sample_selection}).

Switching to the EFF sample, we find that the correlations are sensitive to the chosen radius. The second and third row of Figure~\ref{fig:all_sratio} show the correlations for the isodensity and effective normalised radii, respectively, measured for the same galaxies (see also Table~\ref{table:correlations}). $R_{\rm{50,HI}}/R_{\rm{50,r}}$ shows positive correlations with $M_*$ and $\mu_*$ (opposite to $R_{\rm{iso,HI}}/R_{\rm{iso,r}}$), no significant correlations with sSFR or $\mathrm{NUV}-r$ colour and a weaker positive correlation with $\Delta\,\rm{SFMS}$. The change in the correlations is not due to the smaller EFF sample size as the correlations with $R_{\rm{iso,HI}}/R_{\rm{iso,r}}$ show only small changes for the EFF sample (panels d--f of Figure~\ref{fig:all_sratio} and second row of Table~\ref{table:correlations}). We note that some of the correlation between $R_{\rm{50,HI}}/R_{\rm{50,r}}$ and $\mu_*$ (Equation~\ref{equ:star_surface_density}) is due to the axes being correlated.

This raises the question of what causes the correlations to invert ($M_*$ and $\mu_*$) or be removed (star formation) between isodensity/isophotal and effective radii measured in the H\,\textsc{i} and $r$-band. As discussed in Section~\ref{s-sec:size_mass_relations}, galaxies' effective H\,\textsc{i} and $r$-band radii increase at different rates with H\,\textsc{i}/stellar mass relative to their isodensity/isophotal radii (Figure~\ref{fig:size_mass_residual}, Table~\ref{table:size_mass_coef}). As a result, the effective size relative to the isodensity/isophotal size of the H\,\textsc{i}/$r$-band disc changes with mass. The difference between $R_{\rm{50,HI}}$ and $R_{\rm{iso,HI}}$ ($\Delta R_{\rm{HI}}$) at $M_{\mathrm{HI}}=10^{8}$ vs $10^{10}\,\mathrm{M}_{\odot}$ is $\Delta R_{\rm{HI}}=0.29$ vs 0.21\,dex, respectively. In contrast, in the $r$-band, the relative difference is $\Delta R_{\rm{r}}=0.28$ vs 0.44\,dex (at $M_{\mathrm{*}}=10^{8}$ vs $10^{10}\,\mathrm{M}_{\odot}$, respectively). We conclude that the switch from normalising by the isophotal to effective $r$-band radius is responsible for the inverted/removed correlations with $R_{\rm{50,HI}}/R_{\rm{50,r}}$. Galaxies with higher stellar masses, which tend to have higher $\mu_*$ and lower levels of star formation (Figure~\ref{fig:sample_selection}), have larger $R_{\rm{50,HI}}/R_{\rm{50,r}}$ than $R_{\rm{iso,HI}}/R_{\rm{iso,r}}$.

\begin{figure*}
	\includegraphics[width=\textwidth]{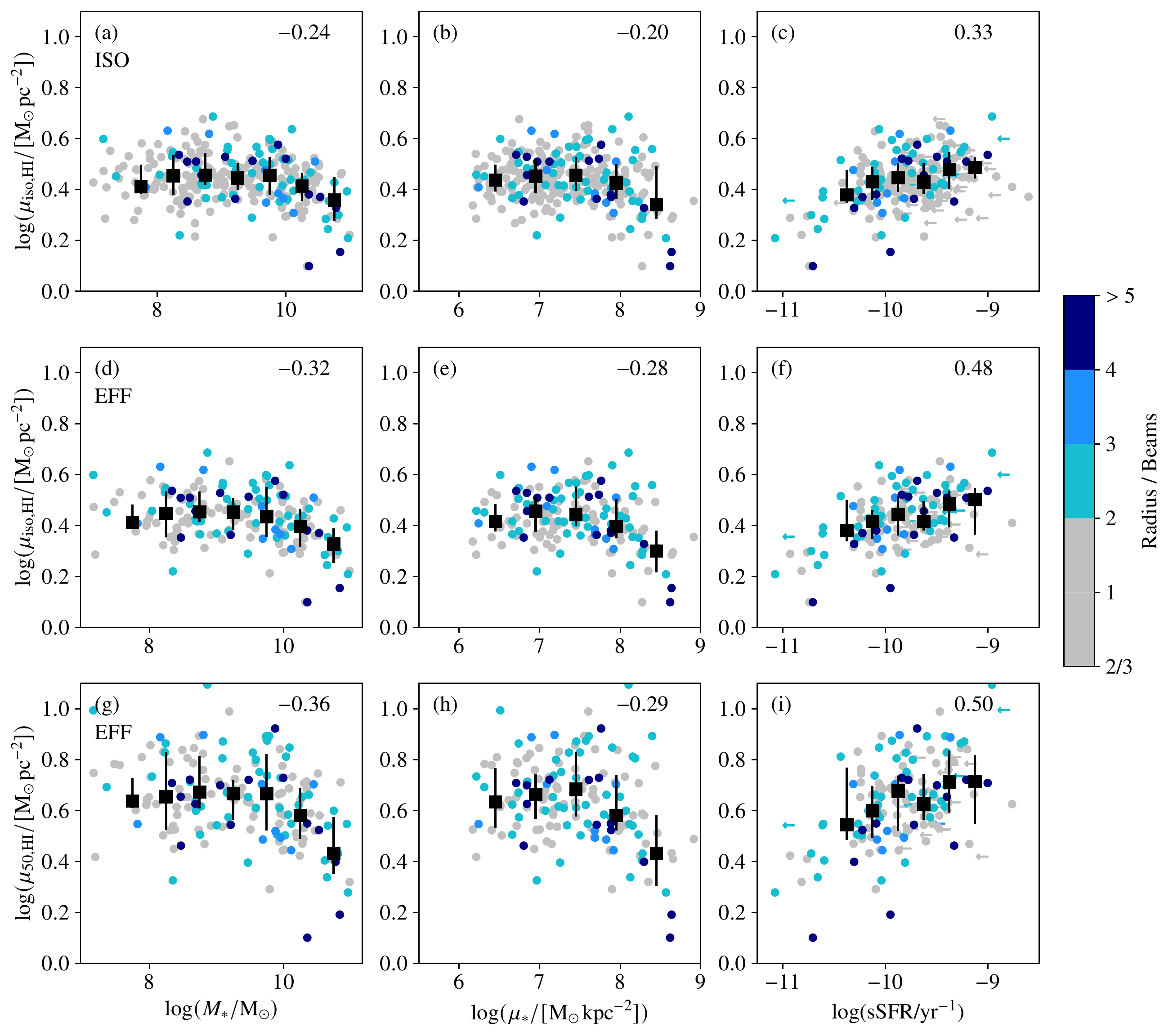}
    \caption{Similar to Figure~\ref{fig:all_sratio}, we now plot the isodensity and effective H\,\textsc{i} surface densities ($\mu_{\rm{iso,HI}}$ and $\mu_{\rm{50,HI}}$) on the $y$-axis and the corresponding Pearson correlation coefficients are tabulated in the last three rows of Table~\ref{table:correlations}.}
    \label{fig:all_muhi}
\end{figure*}

\subsection{The HI Surface Density}
\label{s-sec:hi_mu}

The trends that we find with the normalised H\,\textsc{i} disc size in Section~\ref{s-sec:hi_size} depend on the choice of isophotal or effective $r$-band radius, which are linked to $M_*$ and $\mu_*$. To remove this dependence and investigate the interplay between H\,\textsc{i} structure and galaxy properties directly, we measure the H\,\textsc{i} surface density, $\mu_{\rm{HI}}$, within apertures defined by the effective and isodensity radii. Similar to $R_{\rm{iso,HI}}/R_{\rm{iso,r}}$, we investigate the variation in the isodensity $\mu_{\rm{iso,HI}}$ as a function of $M_*$, $\mu_*$ and sSFR in panels a--c of Figure~\ref{fig:all_muhi} for the ISO sample. We show the Pearson correlation coefficients in the upper right corner of each panel of Figure~\ref{fig:all_muhi} and list them in the fourth row of Table~\ref{table:correlations}. 

We find that galaxies with $M_*>10^{9}\,\mathrm{M}_{\odot}$ and/or $\mu_*>10^{7}\,\mathrm{M}_{\odot}\,\mathrm{kpc}^{-2}$ tend to have marginally lower H\,\textsc{i} surface densities (i.e.\ where the median $\mu_{\rm{iso,HI}}$ values decrease in panels a and b; $\rho=-0.37$ and $-0.32$, respectively), while there is no trend with $\mu_{\rm{iso,HI}}$ for galaxies with lower stellar masses and mass surface densities. Galaxies with higher levels of star formation have higher H\,\textsc{i} surface densities and these correlations strengthen if we only consider galaxies within $\pm0.5$\,dex above/below the SFMS ($\rho=0.37$, $-0.35$ and 0.26 for sSFR, $\mathrm{NUV}-r$ colour and $\Delta\,\rm{SFMS}$, respectively). Star formation shows weaker correlation with $\mu_{\rm{iso,HI}}$ than with $R_{\rm{iso,HI}}/R_{\rm{iso,r}}$.  

A possible explanation for the weak correlations of $M_*$, $\mu_*$ and sSFR with $\mu_{\rm{iso,HI}}$ is that this averages the H\,\textsc{i} surface density over the entire H\,\textsc{i} disc, which is not well matched to the optical/star-forming disc (i.e.\ $\langle R_{\rm{iso,HI}}/R_{\rm{iso,r}}\rangle = 2.1$). Hence, we expect to find stronger correlations within the effective radius for $\mu_{\rm{50,HI}}$, which more closely traces the H\,\textsc{i} within the optical disc (i.e.\ $\langle R_{\rm{50,HI}}/R_{\rm{iso,r}}\rangle = 1.2$) and is co-located with recent star formation. We measure the correlations with the effective $\mu_{\rm{50,HI}}$ for the EFF sample (panels g--i of Figure~\ref{fig:all_muhi}). Although we find stronger correlations between $\mu_{\rm{50,HI}}$ and the five galaxy properties (see the sixth row of Table~\ref{table:correlations}), we find similarly strengthened correlations with the isodensity $\mu_{\rm{iso,HI}}$ for the EFF sample (panels d--f of Figure~\ref{fig:all_muhi} and the fifth row of Table~\ref{table:correlations}). Comparing the lower two rows also highlights that $\mu_{\rm{50,HI}}$ has nearly twice the dynamic range of $\mu_{\rm{iso,HI}}$ (i.e.\ $\mu_{\rm{50,HI,max}}\lesssim10$ vs $\mu_{\rm{iso,HI,max}}\lesssim5\,\mathrm{M}_{\odot}\,\mathrm{pc}^{-2}$). The reduced dynamic range of $\mu_{\rm{iso,HI}}$ is due to the exponential decline of the radial profile beyond $r \gtrsim R_{\mathrm{50,HI}}$. 

The strengthened $\mu_{\rm{iso,HI}}$ correlations for the EFF sample suggest that galaxies with poor spatial resolution (i.e.\ $R_{\mathrm{50,HI}}<20$\,arcsec) are responsible for weakening the correlations in the ISO sample. This also explains the stronger correlations that we find above $M_*>10^{9}\,\mathrm{M}_{\odot}$ and $\mu_*>10^{7}\,\mathrm{M}_{\odot}\,\mathrm{kpc}^{-2}$ in the ISO sample. A larger fraction of these more massive galaxies are resolved by $>4$ beams (e.g.\ $\sim5\%$ vs $\sim30\%$ of galaxies with $M_*<10^{9}\,\mathrm{M}_{\odot}$ and $M_*>10^{9}\,\mathrm{M}_{\odot}$, respectively). Hence, compared to the normalised H\,\textsc{i} disc size, correlations with the H\,\textsc{i} surface density are more sensitive to the spatial resolution of the observations (i.e.\ $\mu_{\rm{HI}}$ is less well constrained in marginally resolved sources).

\section{Discussion}
\label{sec:discussion}

Our results show that galaxies with more extended H\,\textsc{i} discs relative to their isophotal optical discs and higher H\,\textsc{i} surface densities (measured within the isodensity H\,\textsc{i} radius) tend to be more star forming. We find weaker correlations with stellar masses and stellar mass surface densities, which shows that galaxies with higher $M_*$ and/or $\mu_*$ tend to have less extended H\,\textsc{i} discs and lower H\,\textsc{i} surface densities. Limiting our sample to better spatially resolved galaxies strengthens the correlations with the H\,\textsc{i} surface density and we find similar results if we measure the H\,\textsc{i} surface density within the effective H\,\textsc{i} radius. We find that normalising the H\,\textsc{i} disc radius by the optical effective radius inverts or removes correlations compared to using the optical isophotal radius: galaxies with higher $M_*$ and/or $\mu_*$ tend to have more extended H\,\textsc{i} discs and there are no correlations with star formation.

There are different possible pathways to introduce the diversity in the inner H\,\textsc{i} surface densities and the relative sizes of the H\,\textsc{i} discs that we find. Variations may be due to physical processes acting within a galaxy (e.g.\ conversion from H\,\textsc{i} to H$_2$ and subsequent star formation, inflow of gas towards the inner star forming disc, ejection of gas by internal feedback), interactions between galaxies and the environment (e.g\ galaxy mergers, \citeauthor{Toomre1972} \citeyear{Toomre1972}; tidal stripping, \citeauthor{Gunn1972} \citeyear{Gunn1972}; or ram pressure stripping, \citeauthor{Moore1996} \citeyear{Moore1996}, \citeauthor{Moore1999} \citeyear{Moore1999}; see also \citeauthor{Cortese2021} \citeyear{Cortese2021} for a recent review) or may be encoded into the galaxy as it forms (i.e.\ during the initial collapse of gas to form a disc if the gas is more evenly distributed or centrally concentrated). We discuss physical processes that may be responsible for the correlations that we find and how our results compare to previous work.

Several studies in the literature find links between variations of galaxies' H\,\textsc{i} radial profiles and stellar structure quantified by morphological type \citep[e.g.][]{Broeils1997, Cayatte1994, Swaters2002, Noordermeer2005} and stellar properties \citep[e.g.][]{Wang2013, Wang2014}. In particular, \cite{Wang2013} find stronger negative correlations between $R_{\rm{iso,HI}}/R_{\rm{iso,r}}$ and $M_*$ and $\mu_*$ ($\rho=-0.44$ and $-0.31$, respectively) in the Bluedisk galaxy sample (driven by low gas fraction galaxies), than we find in this work. Our results indicate that galaxies' stellar structure (i.e.\ $M_*$ and $\mu_*$; traced by emission in the $g$- and $r$-band) plays a smaller role in regulating how H\,\textsc{i} is distributed than suggested by previous studies. 

Contributing to our differing results are the observation bias to preferentially detect gas-rich galaxies (Section~\ref{sec:sample_selection} and Figure~\ref{fig:sample_selection}) and the limited spatial resolution. Compared to previous work, which used high spatial resolution observations, $\sim75\%$ of our galaxies are marginally resolved (e.g.\ $\gtrsim10$ vs $<4$ beams across the H\,\textsc{i} disc, respectively). Hence, we are insensitive to the impact from the centre of the stellar/star-forming disc in our poorly resolved galaxies. Our results are similar to the findings based on estimated $\mu_{\rm{HI}}$ within the optical disc ($R_{90}$) and $R_{\rm{iso,HI}}/R_{\rm{iso,r}}$ for xGASS disc galaxies (i.e.\ no spatial information is present in their H\,\textsc{i} data) from \cite{Wang2020} and \cite{Pan2021}, respectively. 

As expected, our results show that the H\,\textsc{i} structural parameters are more sensitive to the star formation quantities sSFR and $\mathrm{NUV}-r$ colour than to $M_*$ and $\mu_*$ since H\,\textsc{i} is the raw fuel for future star formation and H\,\textsc{i} and SFR surface densities are found to be correlated (e.g.\ \citeauthor{Bigiel2010} \citeyear{Bigiel2010}; \citeauthor{Wang2017} \citeyear{Wang2017}). \cite{Wang2014} find similar results in the Bluedisk survey with larger differences in the median H\,\textsc{i} surface density profiles within $\sim0.5R_{\rm{iso,HI}}$ as a function of star forming properties than with $M_*$ and $\mu_*$. The correlations that we find with $\Delta\,\rm{SFMS}$ are weaker than with sSFR and $\mathrm{NUV}-r$ colour (i.e.\ when controlling for stellar mass), but are in general agreement with the results of \cite{Wang2020}. 

We find that the H\,\textsc{i} structural parameters show a weaker dependence on the stellar and star formation quantities ($M_*$, $\mu_*$, sSFR and $\mathrm{NUV}-r$) than the H\,\textsc{i} gas fraction ($M_{\mathrm{HI}}/M_*$) for a control xGASS sample ($\rho=-0.83$, $-0.73$, $0.69$ and $-0.78$, respectively, where we have taken a subsample from xGASS with $M_{\mathrm{HI}}/M_*$ above the xGASS medians and sSFR $>2\sigma$ below the xGASS SFMS from \citeauthor{Janowiecki2020} \citeyear{Janowiecki2020}, i.e.\ galaxies comparable to our WALLABY sample). We find that the median $M_{\mathrm{HI}}/M_*$ decreases by $\sim1$\,dex across 2\,dex in stellar mass from $M_*=10^{9}$--$10^{11}\,M_{\odot}$. In contrast, over $\sim3$\,dex from $M_*=10^{8}$--$10^{11}\,M_{\odot}$, $\mu_{\rm{iso,HI}}$ and $R_{\rm{iso,HI}}/R_{\rm{iso,r}}$ both vary by $\sim0.2$\,dex. This indicates that, while the relative amount of H\,\textsc{i} decreases as $M_*$ and $\mu_*$ increase and the level of star formation decreases, the H\,\textsc{i} surface density and the H\,\textsc{i} disc size relative to that of the optical disc vary little for the average galaxy. However, the scatter ($\sim0.1$\,dex) in the H\,\textsc{i} structural parameters indicates that there is significant galaxy to galaxy variation in $\mu_{\rm{iso,HI}}$ and $R_{\rm{iso,HI}}/R_{\rm{iso,r}}$. Some of this scatter is due to effects of low spatial resolution for many galaxies in our sample (Section~\ref{s-sec:hi_data}), however there are physical mechanisms that are also likely contributing, which we discuss below.

Our sample contains galaxies in a range of environments: field galaxies and galaxies in and around the Hydra cluster and NGC\,4636 galaxy group. \cite{Reynolds2022} show that, compared to field galaxies, Hydra cluster members have smaller normalised H\,\textsc{i} discs (i.e.\ median H\,\textsc{i} to optical $r$-band diameter ratios of 3.5 vs 1.9) and conclude that this H\,\textsc{i} disc truncation is likely due to ram pressure stripping, which \cite{Wang2021} identify as impacting many galaxies detected by WALLABY in the Hydra cluster. Hence, some of the scatter present in $R_{\rm{iso,HI}}/R_{\rm{iso,r}}$ is due to environmental processes \citep[e.g.\ ram pressure stripping, which is also observed in group environments,][]{Rasmussen2006, Westmeier2011, Rasmussen2012, Cortese2021}. The scatter in $\mu_{\rm{iso,HI}}$ is similar to that of $R_{\rm{iso,HI}}/R_{\rm{iso,r}}$ ($\sim0.1$\,dex). Ram pressure stripping likely has less impact on $\mu_{\rm{iso,HI}}$ and contributes less to the observed scatter as H\,\textsc{i} is preferentially stripped from a galaxy's outskirts and, for the galaxy sample detected by WALLABY, does not impact star formation or H\,\textsc{i} within the optical disc \citep[e.g.\ as shown by][]{Reynolds2022}. Hence ram pressure stripping is likely to be a second-order effect. Galaxy mergers and tidal stripping are unlikely to affect many galaxies in our sample as we see no obvious signs of disturbances in the optical images and we exclude systems with multiple optical counterparts contained within the H\,\textsc{i} envelope of each detection (Section~\ref{sec:sample_selection}).

H\,\textsc{i} does not form stars directly, but via conversion to H$_2$, which likely contributes to several of our results. Assuming no interactions, mergers or gas replenishment, the conversion of H\,\textsc{i} to H$_2$ in the galaxies' inner regions has little effect on $R_{\rm{iso,HI}}/R_{\rm{iso,r}}$, which can explain the stronger correlations that we find with $R_{\rm{iso,HI}}/R_{\rm{iso,r}}$ compared to $\mu_{\rm{HI}}$ with star formation. The stellar disc grows and the H\,\textsc{i} surface density decreases as the gas is consumed, which leads to lowering the level of star formation. If there is no further accretion of H\,\textsc{i}, then the relative size of the H\,\textsc{i} disc decreases.

The conversion of H\,\textsc{i} to H$_2$ likely has a greater impact on our results with the H\,\textsc{i} surface density as $\mu_{\rm{HI}}$ provides an incomplete picture of the gas within the disc (e.g.\ \citeauthor{Lee2022} \citeyear{Lee2022} detect CO, a proxy for H$_2$, in a sample of 10 galaxies in the NGC\,4636 group above $M_*>10^{9}\,\mathrm{M}_{\odot}$). To first order, below $M_*<10^{9.5}\,\mathrm{M}_{\odot}$ and $\mu_*<10^{7.5}\,\mathrm{M}_{\odot}\,\mathrm{kpc}^{-2}$ we find that $\mu_{\rm{HI}}$ does not correlate with either $M_*$ or $\mu_*$. However, the significant scatter in $\mu_{\rm{iso,HI}}$ and $\mu_{\rm{50,HI}}$ (0.1 and 0.2\,dex, respectively) shows that $\mu_{\rm{HI}}$ is not constant and the conversion of H\,\textsc{i} to H$_2$ likely plays a role, but future studies are needed. 

We find more significant trends of decreasing $\mu_{\rm{iso,HI}}$ and $\mu_{\rm{50,HI}}$ at higher stellar masses. This may be due to higher  stellar mass surface densities as we find similar decreases in $\mu_{\rm{iso,HI}}$ and $\mu_{\rm{50,HI}}$ for galaxies with high $\mu_*$. The change in $\mu_{\rm{iso,HI}}$ and $\mu_{\rm{50,HI}}$ may also provide some information on quenching mechanisms acting in these higher mass galaxies (e.g.\ stellar winds, supernova feedback, H\,\textsc{i} remaining stable against fragmentation and collapse into H$_2$, low star formation efficiency and/or gas removal by stripping by the environment; see review by \citeauthor{Cortese2021} \citeyear{Cortese2021}). Our analysis shows how there is more diversity in the H\,\textsc{i} properties than what may have been thought. The full WALLABY survey will provide further insights into what physical processes are affecting the H\,\textsc{i} surface density.

\section{Conclusions}
\label{sec:conclusion}

We have investigated how variations in H\,\textsc{i} structural parameters of nearby galaxies correlate with $M_*$, $\mu_*$ and properties measuring star formation using uniformly measured spatially resolved H\,\textsc{i} data from WALLABY and multi-wavelength imaging data from PanSTARRS, \textit{GALEX} and WISE. We quantify the H\,\textsc{i} structure using the H\,\textsc{i} surface density, $\mu_{\rm{HI}}$, and H\,\textsc{i} disc size normalised by the optical disc, $R_{\rm{HI}}/R_{\rm{r}}$, measured using effective and isodensity (at 1\,$\mathrm{M}_{\odot}\,\mathrm{pc}^{-2}$) H\,\textsc{i} radii. Our main results are as follows:

\begin{itemize}
    \item The H\,\textsc{i} size-mass relations derived using isodensity and effective radii have similar slopes, which indicates that both radii increase at a similar rate with the H\,\textsc{i} mass. However, the effective size-mass relation has a larger scatter than the isodensity relation, which we attribute to variations in the inner H\,\textsc{i} surface density (e.g.\ within the effective radius, $\mu_{\mathrm{50,HI}}\sim1$--10\,$\mathrm{M}_{\odot}\,\mathrm{pc}^{-2}$).
    
    \item Galaxies with higher stellar masses and stellar surface densities tend to have lower H\,\textsc{i} surface densities and less extended H\,\textsc{i} discs relative to their optical extent. We find weak correlations between the isodensity H\,\textsc{i} quantities ($\mu_{\rm{iso,HI}}$ and $R_{\rm{iso,HI}}/R_{\rm{iso,r}}$) and stellar quantities ($M_*$ and $\mu_*$). This may indicate that these stellar structure quantities plays a minor role in regulating the H\,\textsc{i} surface density and disc extent. However, the majority of our sample are marginally resolved, which limits our ability to probe H\,\textsc{i} located within the stellar disc and contributes to weakening the correlations. We find stronger correlations if we limit our sample to galaxies with $M_*>10^{9}\,M_{\odot}$ and $\mu_*>10^{7}\,\mathrm{M}_{\odot}\,\mathrm{kpc}^{-2}$, which includes a higher fraction of galaxies spatially resolved by $>4$ beams.
    
    \item Galaxies with higher levels of star formation (higher sSFR, lower $\mathrm{NUV}-r$ colour and above the main sequence) tend to have more extended H\,\textsc{i} discs and higher $\mu_{\rm{iso,HI}}$: the isodensity $\mu_{\rm{iso,HI}}$ and $R_{\rm{iso,HI}}/R_{\rm{iso,r}}$ are more strongly correlated with the star-forming quantities (sSFR, $\mathrm{NUV}-r$ colour, $\Delta\,\rm{SFMS}$). 
    
    \item The two H\,\textsc{i} structural parameters have significant scatter, which highlights the diversity of the H\,\textsc{i} properties when controlling for $M_*$, $\mu_*$ or star formation. Limited spatial resolution is partially responsible, but environmental processes (e.g.\ ram pressure stripping) and the conversion from H\,\textsc{i} to H$_2$ likely also contribute to observed scatter.
    
    \item Care must be taken when defining the optical radius as it affects the correlations with the normalised H\,\textsc{i} disc size. The negative correlations with $M_*$ and $\mu_*$ using isophotal radii become positive and the correlations with sSFR and $\mathrm{NUV}-r$ colour using isophotal radii are removed if the H\,\textsc{i} radius is normalised by the effective $r$-band radius. This is due to the growth of the effective optical and H\,\textsc{i} radii relative to the isophotal/isodensity radii in more massive galaxies.
\end{itemize}

Our results show qualitative agreement with previous studies that used either smaller galaxy samples of spatially resolved H\,\textsc{i} observations or estimated H\,\textsc{i} sizes and average surface densities from single dish observations and the H\,\textsc{i} mass-size relation. This work, although limited in spatial resolution, is the first step in providing large, statistical samples of galaxies with spatially resolved H\,\textsc{i} from which H\,\textsc{i} surface densities and disc sizes can be directly measured and compared with other galaxy quantities. With the full WALLABY survey due to start in late 2022, the number of galaxies detected by WALLABY will begin to grow rapidly and WALLABY will produce truly statistically significant galaxy samples with spatially resolved H\,\textsc{i} data.

\begin{acknowledgement}
We thank the anonymous referee for their comments.

This research was conducted by the Australian Research Council Centre of Excellence for All Sky Astrophysics in 3 Dimensions (ASTRO 3D), through project number CE170100013. 

The Australian SKA Pathfinder is part of the Australia Telescope National Facility which is managed by the Commonwealth Scientific and Industrial Research Organisation (CSIRO). Operation of ASKAP is funded by the Australian Government with support from the National Collaborative Research Infrastructure Strategy. ASKAP uses the resources of the Pawsey Supercomputing Centre. Establishment of ASKAP, the Murchison Radio-astronomy Observatory (MRO) and the Pawsey Supercomputing Centre are initiatives of the Australian Government, with support from the Government of Western Australia and the Science and Industry Endowment Fund. We acknowledge the Wajarri Yamatji as the traditional owners of the Observatory site. We also thank the MRO site staff. This paper includes archived data obtained through the CSIRO ASKAP Science Data Archive, CASDA (\url{http://data.csiro.au}). 

L.C. acknowledges support from the Australian Research Council Discovery Project and Future Fellowship funding schemes (DP210100337, FT180100066).

P.K. acknowledges financial support by the German Federal Ministry of Education and Research (BMBF) under grant 05A17PC2 (Verbundprojekt D-MeerKAT-II).

J.M.vdH. acknowledges funding from the Europeaní Research Council under the European Union’s Seventh Framework Programme (FP/2007-2013)/ERC Grant Agreement No. 291531 (‘HIStoryNU’).

D.A.L. and K.S. acknowledge support from the Natural Science and Engineering Research Council of Canada.

A.B. acknowledges support from the Centre National d'Etudes Spatiales (CNES), France.

F.B. would like to acknowledge funding from the European Research Council (ERC) under the European Union’s Horizon 2020 research and innovation programme (grant agreement No.726384/Empire).

The Pan-STARRS1 Surveys (PS1) and the PS1 public science archive have been made possible through contributions by the Institute for Astronomy, the University of Hawaii, the Pan-STARRS Project Office, the Max-Planck Society and its participating institutes, the Max Planck Institute for Astronomy, Heidelberg and the Max Planck Institute for Extraterrestrial Physics, Garching, The Johns Hopkins University, Durham University, the University of Edinburgh, the Queen's University Belfast, the Harvard-Smithsonian Center for Astrophysics, the Las Cumbres Observatory Global Telescope Network Incorporated, the National Central University of Taiwan, the Space Telescope Science Institute, the National Aeronautics and Space Administration under Grant No. NNX08AR22G issued through the Planetary Science Division of the NASA Science Mission Directorate, the National Science Foundation Grant No. AST-1238877, the University of Maryland, Eotvos Lorand University (ELTE), the Los Alamos National Laboratory, and the Gordon and Betty Moore Foundation.

This publication makes use of data products from the Wide-field Infrared Survey Explorer (WISE), which is a joint project of the University of California, Los Angeles, and the Jet Propulsion Laboratory/California Institute of Technology, funded by the National Aeronautics and Space Administration.

This work is based in part on observations made with the Galaxy Evolution Explorer (\textit{GALEX}). \textit{GALEX} is a NASA Small Explorer, whose mission was developed in cooperation with the Centre National d'Etudes Spatiales (CNES) of France and the Korean Ministry of Science and Technology. \textit{GALEX} is operated for NASA by the California Institute of Technology under NASA contract NAS5-98034.
\end{acknowledgement}

%% file: appendices.tex
\section{Beam Resolution}
\label{appendix:beam_resolution}

As discussed in Section~\ref{sec:sample_selection}, the majority of our galaxy sample is poorly resolved, with a spatial resolution of $<2$ beams along the semi-major axis. We investigate how the spatial resolution affects the measured isodensity and effective radii and average surface densities by convolving the moment 0 maps of galaxies resolved by $>2$ beams to a spatial resolution equivalent to 2 beams. We achieve this by scaling the ASKAP 30\,arcsec synthesised beam by the number of beams by which each galaxy's isodensity and effective radii are resolved, and then convolving the moment 0 map with a 2 dimensional Gaussian with a standard deviation set by this scaled synthesised beam. We then follow the exact same procedure described in Section~\ref{s-sec:hi_data} and in \cite{Reynolds2022} to measure isodensity (effective) radii and surface densities from the moment 0 maps convolved to a spatial resolution of 2 beams along the isodensity (effective) diameter.

\begin{figure*}[ht]
	\includegraphics[width=\textwidth]{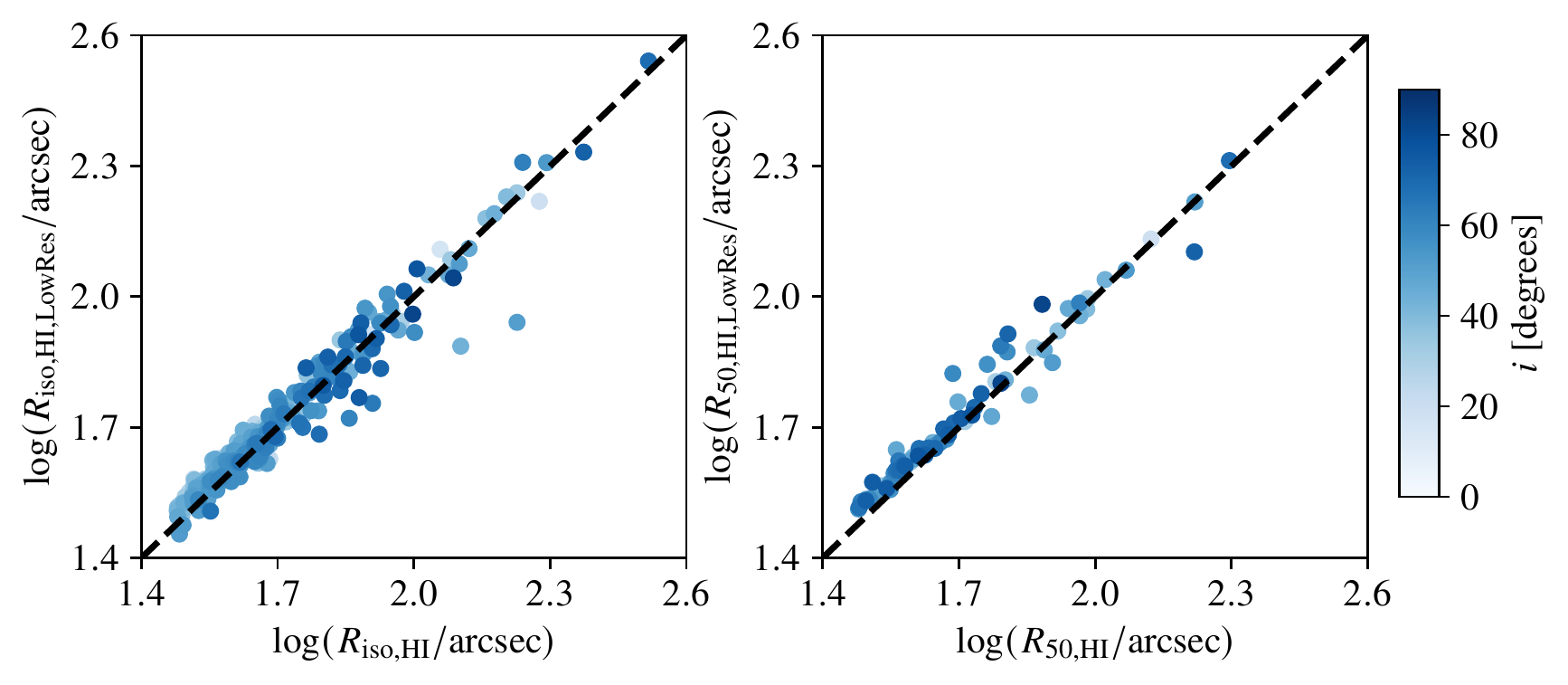}
    \caption{The ISO (left) and EFF (right) radii measured from moment 0 maps convolved to a resolution of 2 beams along the major axis plotted against the radii measured from the full resolution moment 0 maps. The black dash line indicates the 1 to 1 line. The colour bar indicates the inclination of each galaxy as measured from the full resolution moment 0 maps.}
    \label{fig:convolved_parameters}
\end{figure*}

In Figure~\ref{fig:convolved_parameters} we compare the radii measured from these low resolution maps with their original values, and colour the points by the inclination angle measured from the full resolution maps. In both ISO (left) and EFF (right) cases, the data points scatter around the 1 to 1 relation (black dashed line) but tend to lie slightly above it, indicating that radii measured from low resolution maps are somewhat larger. This is a result of the convolution smoothing out and extending the H\,\textsc{i} emission in the maps to larger radii. Correspondingly, we find that the surface densities measured from the low resolution maps slightly decrease (marginally more so for the ISO surface densities). These differences are quantified in Table~\ref{table:beam_resolution}, where we tabulate the medians, 20$^{th}$ and 80$^{th}$ percentiles of the ratios between the radii and surface densities measured from the low and full resolution maps. In all cases, the quantities measured from the low resolution moment 0 maps are within 5-10\% of their original values. Hence, we conclude that the poorly resolved ($<2$ beams) galaxies in our sample have marginally larger radii and lower surface densities than would be measured from higher resolution observations.

\begin{table}
	\centering
	\caption{Median, 20$^{th}$ and 80$^{th}$ percentiles for the distributions in ratios of measured radii and surface densities measured from low and full resolution moment 0 maps.}
	\label{table:beam_resolution}
	\begin{tabular}{lccr}
	    \hline
		& Median & 20$^{th}$ & 80$^{th}$ \\
		\hline
		$R_{\rm{iso,HI,LowRes}}/R_{\rm{iso,HI}}$     & 1.04   & 0.98      & 1.09 \\
		$R_{\rm{50,HI,LowRes}}/R_{\rm{50,HI}}$       & 1.06   & 1.02      & 1.10 \\
        $\mu_{\rm{iso,HI,LowRes}}/\mu_{\rm{iso,HI}}$ & 0.95   & 0.88      & 1.02 \\
        $\mu_{\rm{50,HI,LowRes}}/\mu_{\rm{50,HI}}$   & 0.98   & 0.96      & 1.00 \\
		\hline
	\end{tabular}
\end{table}

\section{Supplementary Data}
\label{appendix:data_table}

We provide a supplementary table in \textsc{fits} format containing all measured and derived galaxy properties that we use in this work. We show the first 10 rows of this file in Table~\ref{table:supplementary_data} with descriptions of the table columns provided here.
\begin{itemize}
    \item Name: WALLABY source detection name
    \item Team Release: Internal WALLABY team release identifier 
    \item $D_{\rm{L}}$: Luminosity distance [Mpc]
    \item $\log(M_{\rm{HI}})$: Base-10 logarithm of the H\,\textsc{i} mass [M$_{\odot}$]
    \item $R_{\rm{50,HI}}$: Effective H\,\textsc{i} radius [kpc]
    \item $R_{\rm{iso,HI}}$: $1\,\mathrm{M}_{\odot}\,\mathrm{pc}^{-2}$ isodensity H\,\textsc{i} radius [kpc]
    \item $\mu_{\rm{50,HI}}$: H\,\textsc{i} surface density measured within the effective H\,\textsc{i} radius [$\mathrm{M}_{\odot}\,\mathrm{pc}^{-2}$]
    \item $\mu_{\rm{iso,HI}}$: H\,\textsc{i} surface density measured within the isodensity H\,\textsc{i} radius [$\mathrm{M}_{\odot}\,\mathrm{pc}^{-2}$]
    \item $\log(M_{\rm{*}})$: Base-10 logarithm of the stellar mass [M$_{\odot}$]
    \item $R_{\rm{50,r}}$: Effective $r$-band radius [kpc]
    \item $R_{\rm{iso,r}}$: 25\,mag\,arcsec$^{-2}$ isophotal $r$-band radius [kpc]
    \item $\log(\mu_*)$: Base-10 logarithm of the stellar mass surface density [$\mathrm{M}_{\odot}\,\mathrm{kpc}^{-2}$]
    \item $\log(\mathrm{sSFR})$: Base-10 logarithm of the specific star formation rate [yr$^{-1}$]
    \item $\mathrm{NUV}-r$: $\mathrm{NUV}-r$ colour [mag]
    \item $\Delta\,\rm{SFMS}$: Offset from the star forming main sequence defined from xGASS by \cite{Janowiecki2020} [dex]
    \item flag$_{\rm{SFR}}$: Flag indicating if the sSFR is an upper limit. For galaxies detected in both \textit{GALEX} and WISE: flag$_{\rm{SFR}}=0$. For galaxies non-detected (i.e.\ $\mathrm{SNR}<5$) in either \textit{GALEX} or WISE (i.e.\ derived sSFR is an upper limit): flag$_{\rm{SFR}}=1$.
\end{itemize}

\begin{landscape}
\begin{table}
	\centering
	\caption{The measured and derived properties for H\,\textsc{i} detected galaxies. The full table is available as supplementary material.}
	\label{table:supplementary_data}
	\footnotesize
	\begin{tabular}{lccccccccccccccr}
	    \hline
		Name  & Team & $D_{\rm{L}}$ & $\log(M_{\rm{HI}})$ & $R_{\rm{50,HI}}$ & $R_{\rm{iso,HI}}$ & $\mu_{\rm{50,HI}}$ & $\mu_{\rm{iso,HI}}$ & $\log(M_*)$ & $R_{\rm{50,r}}$ & $R_{\rm{iso,r}}$ & $\log(\mu_*)$  & $\log(\mathrm{sSFR})$  & $\mathrm{NUV}-r$   & $\Delta\,\rm{SFMS}$ & flag$_{\rm{SFR}}$\\
		WALLABY & Release & [Mpc] & $[\mathrm{M}_{\odot}]$ & [kpc] & [kpc] & [$\mathrm{M}_{\odot}\,\mathrm{pc}^{-2}$] & [$\mathrm{M}_{\odot}\,\mathrm{pc}^{-2}$] & $[\mathrm{M}_{\odot}]$ & [kpc] & [kpc] & $[\mathrm{M}_{\odot}\,\mathrm{kpc}^{-2}]$ & $[\mathrm{yr}^{-1}]$ & [mag] & [dex] &  \\
		\hline
        J100342-270137 & Hydra TR2 & 14.17 & 9.37 & 11.35 & 16.25 & 2.9 & 2.25 & 8.47 & 2.73 & 6.41 & 6.8 & 0.82 & $-9.33$ & 0.31 & 0\\
        J100351-263707 & Hydra TR2 & 12.93 & 8.6 & 2.29 & 4.98 & 9.86 & 3.97 & 7.17 & 0.85 & 1.51 & 6.51 & 0.47 & $-8.85$ & 0.34 & 1\\
        J100351-273417 & Hydra TR2 & 41.52 & 9.84 & 17.56 & 26.62 & 3.54 & 2.4 & 10.35 & 6.87 & 15.63 & 7.88 & 2.24 & $-10.09$ & 0.2 & 0\\
        J100634-295615 & Hydra TR2 & 16.53 & 8.85 & 2.84 & 6.26 & 12.44 & 4.86 & 8.87 & 0.96 & 2.93 & 8.1 & 1.33 & $-8.96$ & 0.82 & 0\\
        J100656-251731 & Hydra TR2 & 43.04 & 9.01 & 5.36 & 10.28 & 4.5 & 2.7 & 8.57 & 3.39 & 6.24 & 6.71 & 1.28 & $-9.52$ & 0.16 & 0\\
        J100707-262300 & Hydra TR2 & 127.37 & 10.02 & 15.94 & 32.89 & 5.41 & 2.86 & 10.14 & 4.36 & 13.39 & 8.06 & 2.02 & $-9.76$ & 0.45 & 0\\
        J100713-262336 & Hydra TR2 & 68.69 & 9.5 & 9.28 & 17.7 & 4.79 & 2.82 & 8.74 & 2.74 & 6.35 & 7.07 & 0.9 & $-9.55$ & 0.18 & 1\\
        J100720-262426 & Hydra TR2 & 254.82 & 10.25 & 20.26 & 43.27 & 4.31 & 2.57 & 10.24 & 10.35 & 24.28 & 7.41 & 2.25 & $-9.94$ & 0.31 & 0\\
        J100752-250626 & Hydra TR2 & 303.6 & 10.42 & 20.31 & 49.66 & 5.39 & 3.02 & 10.31 & 8.76 & 23.35 & 7.63 & 1.76 & $-9.8$ & 0.48 & 0\\
        J100808-260942 & Hydra TR2 & 45.8 & 9.1 & 4.83 & 10.6 & 6.4 & 3.38 & 8.57 & 2.93 & 5.56 & 6.84 & 1.61 & $-9.67$ & 0.0 & 1\\
		\hline
	\end{tabular}
\end{table}
\end{landscape}